\documentclass[11pt]{article}

\usepackage[a4paper,margin=1in]{geometry}
\usepackage{amsmath, amssymb, amsthm, bm}
\usepackage{graphicx, subcaption}
\usepackage{booktabs, multirow, tabularx, threeparttable}
\usepackage{natbib}
\usepackage{setspace}
\usepackage{hyperref}
\usepackage{enumerate}
\usepackage{xcolor}
\usepackage{float}
\usepackage{authblk}
\usepackage{tikz}
\usetikzlibrary{positioning,arrows.meta,fit,backgrounds}
\usepackage{rotating}
\usepackage{wasysym} 
\usepackage{booktabs}
\usepackage{tabularx}
\usepackage{array}
\usepackage{enumitem}
 
\usepackage{amsthm}
\theoremstyle{plain}
\newtheorem{theorem}{Theorem}[section]
\newtheorem{proposition}[theorem]{Proposition}

\newtheorem{corollary}[theorem]{Corollary}
\theoremstyle{definition}

\newtheorem{definition}[theorem]{Definition}
\theoremstyle{remark}
\newtheorem{remark}{Remark}[section]
\newtheorem{example}{Example}[section]

\usepackage{algorithm}
\usepackage{algorithmicx}
\usepackage[noend]{algpseudocode}   
\usepackage{float}                   
\algrenewcommand{\algorithmicrequire}{\textbf{Input:}}
\algrenewcommand{\algorithmicensure}{\textbf{Output:}}

\title{Model-Induced Hazard Geometry and Functional Representations in Survival Analysis}
\author{
  Xijia Liu\thanks{
    Department of Statistics, Umeå University, Sweden.
    \textbf{Email}: \texttt{xijia.liu@umu.se}
  }
}

\date{}

\begin{document}
    \maketitle

\begin{abstract}
Individualized survival predictions are often summarized by scalar risk scores or event probabilities at selected time horizons, although clinically relevant risk may also differ in its temporal organization.
We consider finite-dimensional representations of individual hazard trajectories through interpretable hazard functionals, such as peak timing, cumulative hazard over prespecified intervals, and persistence of elevated risk.

A central issue is that the richness of such representations is constrained by the hazard geometry supported by the underlying survival model.
We formalize this through model-induced tangent spaces and functional ranks, derive an exact decomposition of model-supported variation into directions retained and lost by a chosen representation, and introduce a model-relative notion of local completeness.
We further distinguish intrinsic model capacity from the variation actually activated by covariates.

These results motivate a structured extension of AFT time scaling through a positive-affine deformation on the log-time scale, yielding a power-warp mechanism with an additional temporal deformation direction.
Combined with a log-logistic reference hazard, this gives a parsimonious likelihood-based working specification for which peak timing and peak height provide complete coordinates over the unimodal regime.
We also establish plug-in recovery and asymptotic inference for hazard functionals.
Simulations and a lung-cancer application illustrate how model-induced restrictions propagate to hazard-functional representations.
\end{abstract}


\section{Introduction}
\label{sec:introduction}

Modern survival analysis increasingly supports individualized characterization of time-to-event risk through flexible models \citep{royston2002flexible,Ishwaran2008RSF,Katzman2018,lee2018deephit}.
Depending on the modelling framework, a fitted model may provide a risk score, a survival curve or a full conditional time-to-event distribution.
In practice, these rich outputs are often summarized by scalar risk scores or event probabilities at selected horizons, which are useful for discrimination and decision-making but provide only a compressed description of temporal risk.

When scientific interest concerns more specific aspects of how risk unfolds over time, features of the hazard trajectory, such as when risk reaches its peak, how intense that peak is, or how long elevated risk persists, provide natural and interpretable targets.
Applied biomedical analyses have accordingly used estimated hazards to characterize temporal patterns of recurrence or mortality and to extract features such as peak timing and intensity \citep{simes1985exploratory,saphner1996annual,ding2009parametric,ardoino2012flexible,lima2023application,yamauchi2022944p,kanematsu2022surveillance}.

These practices motivate a systematic statistical treatment of temporal features derived from hazard trajectories.
We formalize such targets as hazard functionals and collect them into a finite-dimensional representation,
$
    G(h)
    =
    \{g_1(h),\ldots,g_K(h)\}^{\top}.
$
Rather than treating these quantities as isolated descriptive summaries, we study the resulting hazard-functional representation as a statistical object, focusing on its model-induced structure and statistical recoverability.
The representation also provides an interpretable scale for describing how fitted covariate--survival relationships are expressed through temporal features of the hazard trajectory.
This perspective leads to two statistical questions.
First, how much independent hazard variation is supported by the model, activated by covariates, and retained by a chosen functional representation?
Second, under what conditions can the resulting hazard functionals be reliably recovered from estimated hazard trajectories?

Classical survival models already illustrate this structural restriction.
Under proportional hazards, covariate effects act through vertical rescaling of a common hazard and therefore do not alter its temporal organization \citep{cox1972regression}.
AFT models permit temporal rescaling, but all temporal features remain governed by a single time-scale coordinate \citep{kalbfleisch2002statistical}.
Thus, several hazard functionals may vary while still reflecting the same underlying deformation direction.
More flexible parametric, spline-based, transformation, and time-varying-effect models can support richer hazard trajectories \citep{royston2002flexible,crowther2014general,hothorn2018most}, but greater flexibility does not by itself reveal how many independent hazard directions are generated or how much of that variation remains visible through a chosen functional representation.

We formalize this issue through \emph{model-induced hazard geometry}.
Local tangent spaces characterize the independent first-order directions supported by a model-induced hazard family, while temporal rank measures the dimension of temporal variation remaining after pure vertical rescaling is removed.
Functional rank then quantifies how much of this model-supported variation remains visible after mapping the hazard through $G$.
Under a first-order compatibility condition, the local model rank decomposes exactly into variation retained by $G$ and model-supported directions that are locally invisible to the representation.
This leads to a model-relative notion of local completeness.
The framework further distinguishes the variation available from the model from that actually activated through the covariate map.

These results also provide a model-design principle: independent variation in a functional representation can only be realized if the underlying model supplies the corresponding hazard directions.
AFT scaling provides only one temporal deformation direction, motivating an extension from log-time translation to positive-affine log-time deformation.
The resulting power-warp mechanism adds a second non-level temporal direction.
Combined with a log-logistic reference hazard, it yields a parsimonious likelihood-based working specification, which we refer to as the log-logistic power-warp (LLP) specification, with the classical log-logistic AFT model as a nested special case.
For a fixed log-logistic reference shape, peak timing and peak height provide complete coordinates for the corresponding two-dimensional LLP hazard family over its unimodal regime.
The LLP specification is a restricted member of a broader class of multiparameter flexible survival models \citep{burke2020flexible} and serves here as a tractable illustration of geometry-guided model extension.

The second issue is statistical recovery.
We establish conditions under which $G\{\widehat h(\cdot\mid x)\}$ consistently recovers $G\{h(\cdot\mid x)\}$.
Uniform hazard convergence provides a general sufficient condition for continuous functionals, while geometry-specific descriptors require additional local stability.
For smooth parameter-induced representations, asymptotic inference follows from the multivariate delta method, and under nondegenerate parameter uncertainty the rank of the asymptotic functional covariance coincides with the local functional rank.
For the log-logistic power-warp specification, we additionally establish identifiability, standard likelihood-based large-sample properties, and uniform recovery of fitted hazard trajectories.

The remainder of the paper is organized as follows.
Section~2 develops the local theory of hazard geometry, functional rank, and covariate-induced variation.
Section~3 introduces the power-warp deformation and its log-logistic implementation.
Section~4 establishes the hazard-functional recovery results.
Section~5 presents simulation studies, Section~6 gives a real-data illustration, and Section~7 concludes with a discussion.

\section{A Local Theory of Hazard Geometry and Functional Rank}
\label{sec:geometry}

This section addresses the first statistical question raised in the Introduction: how many independent dimensions of risk-pattern variation can a survival model support when its fitted hazards are summarized through multiple functionals?
We first define finite-dimensional hazard-functional representations and then characterize the local variation generated by model-induced hazard families.
The resulting rank framework separates the number of reported summaries from the number of independent representational directions actually supported by the model.

\subsection{Hazard Functional Representations}
\label{subsec:hazard_functional_representations}

A fitted hazard trajectory contains information about how risk evolves over time.
Rather than reporting the entire trajectory, selected aspects can be summarized through scalar functionals.
For example, whenever the hazard is integrable on $[0,\tau]$, the event probability at a fixed horizon $\tau$ can be written as $R_\tau(h) = 1-\exp\left\{-\int_0^\tau h(u)\,du\right\}$.
Other functionals may describe temporal features such as peak timing, peak intensity, high-risk duration, or the allocation of hazard across clinically relevant time windows.
This motivates treating several such quantities jointly rather than restricting attention to a single scalar summary.

For the local geometric analysis below, let $\mathcal H$ denote a class of strictly positive continuously differentiable hazard functions on a compact interval $[\varepsilon,T_0]$,
\[
    \mathcal H
    =
    \left\{
        h:[\varepsilon,T_0]\to\mathbb R_{>0}:
        h\in C^1([\varepsilon,T_0])
    \right\},
\]
where $\varepsilon>0$ avoids technical complications at the origin.
We regard $\mathcal H$ as a subset of $L^2([\varepsilon,T_0])$ when defining tangent-space geometry below.
The $L^2$ structure is used for this geometric analysis only; statistical recovery in Section~\ref{sec:functional_recovery} is formulated under a separate supremum-norm topology.

\begin{definition}[Hazard functionals and functional representations]
Let $\mathcal H$ be a class of admissible hazard functions.
A \emph{hazard functional} is a map $g:\mathcal H\to\mathbb R$, where $g(h)$ summarizes a selected aspect of the hazard trajectory.
A \emph{finite-dimensional hazard-functional representation} is a map $G:\mathcal H\to\mathbb R^K$, $G(h)=\bigl(g_1(h),\ldots,g_K(h)\bigr)^\top$, where each $g_k$ is a hazard functional.
Its attainable representation set is
$
    \mathcal R_G
    =
    G(\mathcal H)
    =
    \left\{
        G(h):h\in\mathcal H
    \right\}
    \subseteq\mathbb R^K.
$
\end{definition}

\begin{remark}[Regularity of geometry-specific functionals]
\label{rem:functional_regularity}
The hazard class $\mathcal H$ is not globally restricted to a particular shape.
Some geometry-specific functionals, however, require additional local smoothness
or nondegeneracy conditions for the first-order analysis developed below.
For example, peak-based functionals are considered locally at sufficiently smooth
hazards with a stable interior peak, whereas crossing-based functionals require
stable transversal crossings.
These are functional-specific local conditions rather than global restrictions
on $\mathcal H$; their role is formalized in Section~\ref{subsec:functional_rank_bounds}.
\end{remark}

The representation $G$ is analyst chosen and therefore depends on the scientific questions of interest.
Its $K$ components may target different aspects of risk, including cumulative burden, timing, intensity, persistence, or late-risk behaviour.
Importantly, however, $K$ is only the number of reported summaries; it does not imply that these summaries provide $K$ independent dimensions of variation.
Different functionals may be algebraically or structurally related over the class of hazards under consideration.
The hazard functionals used in the numerical and applied analyses are collected in Appendix~\ref{app:list_hazard_functionals}, where their definitions are considered in relation to the regularity distinctions developed in Section~\ref{subsec:functional_rank_bounds}.

In practice, moreover, individual hazards are not arbitrary elements of $\mathcal H$ but are generated or estimated through a survival model.
The model therefore restricts the subset of hazard trajectories that can occur across covariate profiles, and these restrictions are inherited by $G(h)$.
Accordingly, the effective dimension of a hazard-functional representation depends jointly on the chosen map $G$ and on the variation permitted by the underlying survival model.
The following subsections formalize this model-side restriction through hazard orbits, local tangent spaces, and model-induced ranks.

\subsection{Transformation-Induced Hazard Orbits}
\label{subsec:hg_orbits}

The hazard-functional representation introduced above is constrained by the class of hazard trajectories permitted by the underlying survival model.
To characterize this model-side restriction, we consider survival models that generate covariate-specific hazards through transformations of a reference hazard.
Let $h_0\in\mathcal H$ be a reference hazard and let $\{T_z:z\in\Omega\}$ be a family of transformation operators such that $T_z h_0\in\mathcal H$ for all $z\in\Omega$.
The induced hazard family is $h_z=T_z h_0$, where $z$ indexes the deformation applied to the reference trajectory.

\begin{definition}[Hazard orbit]
\label{def:hazard_orbit}
The hazard orbit generated by $h_0$ and $\{T_z:z\in\Omega\}$ is
\[
    \mathcal O_T(h_0)
    =
    \left\{
        T_z h_0:z\in\Omega
    \right\}
    \subseteq\mathcal H.
\]
\end{definition}

\begin{example}[Classical transformation-induced hazard orbits]
\label{ex:classical_hazard_orbits}
Several classical survival models admit this representation.
Under proportional hazards,
\[
    T_\eta^{\mathrm{PH}}h_0(t)
    =
    e^\eta h_0(t),
\]
so covariate variation generates vertical rescaling of the reference hazard.
Under an accelerated failure time model,
\[
    T_\lambda^{\mathrm{AFT}}h_0(t)
    =
    e^{-\lambda}h_0(te^{-\lambda}),
\]
so variation acts through temporal rescaling.
An AFT model with an additional multiplicative level component, as in multiplicative frailty extensions of AFT models \citep{pan2001using,kats2023accelerated}, has
\[
    T_{\lambda,\eta}^{\mathrm{AFT+L}}h_0(t)
    =
    e^\eta e^{-\lambda}h_0(te^{-\lambda}),
\]
and therefore combines temporal rescaling with vertical level variation.
More generally, $z=(z_1,\ldots,z_q)$ may contain several deformation coordinates.
\end{example}

The orbit $\mathcal O_T(h_0)$ describes the hazard trajectories that are attainable under a given transformation mechanism.
For a chosen hazard-functional representation $G:\mathcal H\to\mathbb R^K$, it induces the representation set
\[
    G\{\mathcal O_T(h_0)\}
    =
    \left\{
        G(T_z h_0):z\in\Omega
    \right\}
    \subseteq\mathbb R^K.
\]
Thus, the model restricts not only which hazard trajectories can occur, but also how the reported hazard functionals can vary jointly.
The dimension of this induced variation need not equal either the number of deformation parameters or the number $K$ of reported functionals.
We therefore turn to the local tangent structure of the hazard orbit, which provides a tractable measure of the independent directions of model-supported hazard variation.

\subsection{Local and Temporal Geometric Expressiveness}
\label{subsec:local_temporal_expressiveness}

We now characterize the independent local directions of hazard variation supported by a model.
Fix a deformation state $z_0\in\Omega$.

\begin{definition}[Local hazard tangent space and overall local rank]
\label{def:local_hazard_tangent_rank}
Let $h_z=T_z h_0$, $z=(z_1,\ldots,z_q)\in\Omega\subseteq\mathbb R^q$, and suppose that $z\mapsto h_z$ is differentiable at $z_0$ as an $L^2([\varepsilon,T_0])$-valued map.
For $j=1,\ldots,q$, define
$
    \dot h_j(\cdot;z_0) = \left. \partial h_z/\partial z_j \right|_{z=z_0}.
$
The local hazard tangent space is
$
    \mathcal T_{\mathcal M}(z_0)
    =
    \operatorname{span}
    \left\{
        \dot h_1(\cdot;z_0),\ldots,
        \dot h_q(\cdot;z_0)
    \right\},
$
and the overall local rank is $r_{\mathcal M}(z_0) = \dim \mathcal T_{\mathcal M}(z_0)$.
\end{definition}

Thus, $r_{\mathcal M}(z_0)$ counts the linearly independent first-order hazard deformations that the model can generate locally.
Overall rank does not distinguish the nature of these directions.
In particular, adding a pure vertical level component can increase $r_{\mathcal M}$ without introducing a new temporal pattern of risk.
To isolate temporal deformation, we work on the log-hazard scale.
Let $\mathcal A_T=L^2([\varepsilon,T_0])$, $V=\operatorname{span}\{1\}$, and let $ \pi:\mathcal A_T\to\mathcal A_T/V $ denote the quotient projection.
Because multiplying a hazard by a positive constant adds a constant to its log-hazard, the quotient identifies hazards that differ only by vertical rescaling.

\begin{definition}[Temporal tangent space and temporal rank]
\label{def:temporal_tangent_rank}
Suppose that the log-hazard map $z\mapsto \ell_z=\log h_z$ is differentiable at $z_0$ as an $\mathcal A_T$-valued map.
Let 
$
    \dot\ell_j(\cdot;z_0)
    =
    \left.
    \partial \ell_z/\partial z_j
    \right|_{z=z_0}
$, $j=1,\ldots,q$.
The projected temporal tangent space is
$
    \bar{\mathcal T}_{\mathcal M}(z_0)
    =
    \operatorname{span}
    \left\{
        \pi(\dot\ell_1(\cdot;z_0)),\ldots,
        \pi(\dot\ell_q(\cdot;z_0))
    \right\}
    \subseteq \mathcal A_T/V,
$
and the temporal rank is
$
    r_{\mathrm{temp},\mathcal M}(z_0) = \dim \bar{\mathcal T}_{\mathcal M}(z_0).
$
\end{definition}

The temporal rank therefore counts the independent local deformation directions that remain after pure vertical rescaling has been removed.
The distinction between overall and temporal rank is illustrated by the basic transformation mechanisms introduced above.

\begin{proposition}[Overall and temporal ranks of basic transformation orbits]
\label{prop:overall_temporal_ranks_basic_orbits}
Let $h_0\in\mathcal H$ and write $\ell_0=\log h_0$.
Consider the proportional hazards, AFT, and AFT-plus-level transformations in Example~\ref{ex:classical_hazard_orbits}, evaluated at the identity deformation state.
Assume the non-degeneracy condition, i.e., $t\ell_0'(t)$ is not constant on $[\varepsilon,T_0]$.
Then
\[
    \begin{array}{c|cc}
        \text{Model}
        &
        r_{\mathcal M}
        &
        r_{\mathrm{temp},\mathcal M}
        \\
        \hline
        \mathrm{PH}     & 1 & 0 \\
        \mathrm{AFT}    & 1 & 1 \\
        \mathrm{AFT+L}  & 2 & 1
    \end{array}
\]
at the identity deformation state.
\end{proposition}

Thus, parameter dimension and overall hazard rank do not by themselves determine temporal representational capacity.
The AFT-plus-level model has two independent hazard directions, but only one remains after vertical rescaling is removed.
The proof is given in Appendix~\ref{app:proof_overall_temporal_ranks_basic_orbits}.

\subsection{Model-Induced Functional Rank and Rank Loss}
\label{subsec:functional_rank_bounds}

We now ask how much independent first-order variation can be revealed by scientifically interpretable hazard functionals under a given survival model.
Let $h_z=T_z h_0$ and define $\Psi_G(z)=G(h_z)$, $G:\mathcal H\to\mathbb R^K$.
The central question is whether the variation visible through $G$ can exceed the local hazard variation supported by the model itself.

For the local analysis below, it is useful to distinguish two levels of
functional regularity.
We call a representation \emph{ambient-space regular} at $h_{z_0}$ if
$G$ is Fr\'echet differentiable at that hazard state.
Provided that $z\mapsto h_z$ is differentiable, this implies
$
    D\Psi_G(z_0)
    =
    DG_{h_{z_0}}\circ Dh_{z_0}.
$
A weaker notion is \emph{model-path regularity}, which requires only that the finite-dimensional map $z\mapsto G(h_z)$ be differentiable at $z_0$.
Thus, ambient-space regularity implies model-path regularity, whereas the converse need not hold.
Geometry-specific functionals may be model-path regular without being regular under arbitrary perturbations in the ambient hazard space.
For example, peak location may be differentiable along a sufficiently smooth model family at a stable nondegenerate peak, and threshold-based functionals may be differentiable at stable transversal crossings.
At degenerate configurations, such as tied or flat peaks or tangential crossings, model-path differentiability itself may fail; such points fall outside the ordinary first-order functional-rank analysis considered below.
These regularity properties are local to the functional, model, and hazard state rather than intrinsic labels attached to a functional in isolation.

\begin{definition}[Local $G$-functional rank]
\label{def:local_G_functional_rank}
Suppose that $\Psi_G:\Omega\to\mathbb R^K$ is differentiable at $z_0$.
The local $G$-functional rank of model $\mathcal M$ at $z_0$ is
\[
    r_{G,\mathcal M}(z_0)
    =
    \operatorname{rank}\{D\Psi_G(z_0)\}
    =
    \operatorname{rank}
    \left[
        \left.
        \frac{\partial G(h_z)}
        {\partial z^\top}
        \right|_{z=z_0}
    \right].
\]
\end{definition}

The rank $r_{G,\mathcal M}(z_0)$ is the number of independent first-order representational directions visible through the chosen functionals.
It depends jointly on the model and on $G$.
Thus, a model may generate several independent first-order hazard directions while a particular representation retains only some of them.
Importantly, Definition~\ref{def:local_G_functional_rank} requires only model-path regularity and does not require Fr\'echet differentiability of $G$ with respect to arbitrary perturbations in the ambient hazard space.

For the rank comparison below, we focus on representations whose first-order variation is compatible with the first-order hazard geometry.
For every parameter direction $a\in\mathbb R^q$, suppose that
\begin{equation}
\label{eq:functional_first_order_compatibility}
    Dh_{z_0}[a]=0
    \quad\Longrightarrow\quad
    D\Psi_G(z_0)[a]=0.
\end{equation}
Condition~\eqref{eq:functional_first_order_compatibility} states that a parameter perturbation that is invisible to the hazard at first order must also be invisible to its functional representation at first order.
It therefore places hazard variation and functional variation on the same first-order scale, which is the scale considered by the rank analysis below.
Functionals with first-order sensitivity to higher-order hazard variation may still be informative, but fall outside this first-order comparison.

The compatibility condition holds automatically when $G$ is Fr\'echet differentiable.
It is also automatic for a locally nonredundant deformation parameterization, for which $r_{\mathcal M}(z_0)=q$.
The following proposition provides a concrete example of model-path regularity and first-order compatibility for a geometry-specific representation.

\begin{proposition}[Regularity of peak-based functionals]
\label{prop:peak_functional_regularity}
Suppose that $h_z(t)$ is continuously differentiable in $z$ and twice continuously differentiable in $t$ on a neighbourhood of $\{z_0\}\times[\varepsilon,T_0]$, with the relevant mixed derivatives continuous, and that $h_{z_0}$ has a unique interior nondegenerate maximum $t_p\in(\varepsilon,T_0)$ satisfying $h_{z_0}'(t_p)=0$ and $h_{z_0}''(t_p)<0$. 
Then the peak-based representation
$
    G_{\mathrm{peak}}(h)
    =
    \{t_{\mathrm{peak}}(h),h_{\mathrm{peak}}(h)\}^{\top}
$
is differentiable along the model-induced family at $z_0$ and satisfies the first-order compatibility condition~\eqref{eq:functional_first_order_compatibility}.
\end{proposition}

The proof is given in Appendix~\ref{proof:peak_functional_regularity}.
We now turn to the general relation between model-supported hazard rank and functional rank.

\begin{theorem}[Exact functional rank decomposition and temporal rank bound]
\label{thm:functional_rank_bound}
Let $h_z=T_z h_0$ be differentiable at $z_0$, and suppose that $\Psi_G(z)=G(h_z)$ is differentiable at $z_0$.

\begin{enumerate}
\item[(i)]
If the first-order compatibility condition
\eqref{eq:functional_first_order_compatibility} holds, then the rule
$
    L_{G,\mathcal M}(z_0)\{Dh_{z_0}[a]\}
    =
    D\Psi_G(z_0)[a],
$
$    
    a\in\mathbb R^q,
$
defines a unique linear map
$
    L_{G,\mathcal M}(z_0):
    \mathcal T_{\mathcal M}(z_0)
    \longrightarrow
    \mathbb R^K.
$
The local model rank then admits the exact decomposition
\[
    r_{\mathcal M}(z_0)
    =
    r_{G,\mathcal M}(z_0)
    +
    \dim
    \left\{
        v\in\mathcal T_{\mathcal M}(z_0):
        L_{G,\mathcal M}(z_0)[v]=0
    \right\}.
\]
In particular,
$
    r_{G,\mathcal M}(z_0)
    \leq
    r_{\mathcal M}(z_0).
$
If, in addition, $G$ is Fr\'echet differentiable at $h_{z_0}$, then
$
    L_{G,\mathcal M}(z_0)
    =
    DG_{h_{z_0}}
    \big|_{\mathcal T_{\mathcal M}(z_0)},
$
and the decomposition becomes
\[
    r_{\mathcal M}(z_0)
    =
    r_{G,\mathcal M}(z_0)
    +
    \dim
    \left\{
        \mathcal T_{\mathcal M}(z_0)
        \cap
        \ker(DG_{h_{z_0}})
    \right\}.
\]

\item[(ii)]
If the temporal tangent space $\bar{\mathcal T}_{\mathcal M}(z_0)$ is well defined, $G$ is invariant to positive vertical rescaling, and the induced map on the quotient log-hazard space $\mathcal A_T/\mathcal V$ is Fr\'echet differentiable at $\pi(\ell_{z_0})$, then
$
    r_{G,\mathcal M}(z_0)
    \leq
    r_{\mathrm{temp},\mathcal M}(z_0).
$
\end{enumerate}
\end{theorem}

The proof is given in Appendix~\ref{proof:geom_bound}.
Part~(i) decomposes the local model rank into variation retained by $G$ and model-supported first-order variation that is locally invisible to $G$.
Accordingly, the difference
$
    r_{\mathcal M}(z_0)-r_{G,\mathcal M}(z_0)
$
may be interpreted as the local functional rank loss.
Under the conditions of Theorem~\ref{thm:functional_rank_bound}(i), this loss is exactly
$
    \dim
    \left\{
        v\in\mathcal T_{\mathcal M}(z_0):
        L_{G,\mathcal M}(z_0)[v]=0
    \right\}.
$

\begin{definition}[Local completeness]
\label{def:local_completeness}
Under the conditions of Theorem~\ref{thm:functional_rank_bound}(i), the
representation $G$ is said to be locally complete for $\mathcal M$ at $z_0$ if
$
    r_{G,\mathcal M}(z_0)
    =
    r_{\mathcal M}(z_0).
$
\end{definition}

Local completeness is model-relative.
It requires the representation to distinguish all first-order hazard directions generated by the model, rather than arbitrary perturbations in the surrounding hazard function space.
Thus, a finite-dimensional representation need not characterize an arbitrary hazard trajectory in order to be locally complete for a particular model-induced family.
A concrete example is given in Section~\ref{sec:power_warp}, where peak timing and peak height are shown to provide complete coordinates for the fixed-$\alpha$ log-logistic power-warp hazard family over its unimodal regime.
This distinction is reflected in the dimensional bounds on functional rank.
For a representation consisting of $K$ hazard functionals,
$
    r_{G,\mathcal M}(z_0)
    \leq
    \min\{K,r_{\mathcal M}(z_0)\},
$
and, for vertically invariant representations satisfying the stated quotient-space regularity conditions,
$
    r_{G,\mathcal M}(z_0)
    \leq
    \min\{K,r_{\mathrm{temp},\mathcal M}(z_0)\}.
$
Hence, the number of reported summaries, the number of independent first-order directions retained by those summaries, and the number of directions supported by the model are distinct quantities.
Additional functionals may enrich scientific interpretation, but cannot increase the intrinsic dimension of model-supported functional variation.
We next consider how these quantities enter a regression formulation.

\subsection{Reparameterization Invariance and Covariate-Induced Rank}
\label{subsec:covariate_induced_rank}

The ranks introduced above describe local variation in a model-induced hazard
family.
Two further distinctions are important before moving to a regression
specification.
First, these ranks should not depend on the particular smooth coordinates used
to index the same hazard family.
Second, even when a deformation family supports several independent hazard
directions, the covariate map used in a regression model need not activate all
of them.
We first establish invariance to regular reparameterization.

\begin{proposition}[Reparameterization invariance]
\label{prop:rank_reparameterization}
Suppose that the same local hazard family is parameterized by $z\in\Omega$ and $\vartheta\in\widetilde\Omega$, with
$
    \widetilde h_\vartheta
    =
    h_{\phi(\vartheta)},
$
$
    z_0=\phi(\vartheta_0),
$
where $\phi$ is a $C^1$ local diffeomorphism at $\vartheta_0$.
Then the local hazard tangent spaces generated by the two parameterizations coincide, and hence
$
    \widetilde r_{\mathcal M}(\vartheta_0)
    =
    r_{\mathcal M}(z_0).
$
For any model-path regular representation $G$,
$
    \widetilde r_{G,\mathcal M}(\vartheta_0)
    =
    r_{G,\mathcal M}(z_0).
$
If the temporal tangent spaces are well defined, then also
$
    \widetilde r_{\mathrm{temp},\mathcal M}(\vartheta_0)
    =
    r_{\mathrm{temp},\mathcal M}(z_0).
$
Consequently, the first-order compatibility condition, local functional rank loss, and local completeness are invariant under such reparameterizations.
\end{proposition}

The proof is given in Appendix~\ref{proof:rank_reparameterization}. Thus, the ranks are properties of the local model-induced hazard family rather than artifacts of a particular regular coordinate system.
In particular, introducing additional named parameters does not by itself increase geometric expressiveness if those parameters do not generate new independent hazard directions.

We next distinguish the variation supported by the deformation family from the variation actually activated by covariates.
In a regression formulation, this distinction arises when the deformation parameter $z$ is modeled as a function of covariates.
Let $x\in\mathcal X\subseteq\mathbb R^p$ and suppose that $z=\eta(x)$, where $\eta:\mathcal X\to\Omega$ is differentiable.
The resulting covariate-indexed hazard is $h_{\eta(x)}$.

\begin{definition}[Covariate-induced ranks]
\label{def:covariate_induced_rank}
At $x_0\in\mathcal X$, with $z_0=\eta(x_0)$, define the covariate-induced
hazard tangent space by
$
    \mathcal T_X(x_0)
    =
    \operatorname{Im}
    \left\{
        \left.
        D_x h_{\eta(x)}
        \right|_{x=x_0}
    \right\},
$
with covariate-induced hazard rank
$
    r_X(x_0)
    =
    \dim\{\mathcal T_X(x_0)\}.
$
For a hazard-functional representation $G$, define
$
    \Phi_G(x)
    =
    G\{h_{\eta(x)}\}
    =
    \Psi_G\{\eta(x)\},
$
and, whenever $\Phi_G$ is differentiable at $x_0$, define the covariate-induced $G$-functional rank by
$
    r_{G,X}(x_0)
    =
    \operatorname{rank}\{D\Phi_G(x_0)\}.
$
\end{definition}

The quantities $r_{\mathcal M}$ and $r_{G,\mathcal M}$ describe the local capacity of the deformation family, whereas $r_X$ and $r_{G,X}$ describe the part of that capacity activated through the covariate map.

\begin{proposition}[Covariate-induced rank bounds]
\label{prop:covariate_induced_rank_bound}
Suppose that $h_z$ and $\Psi_G(z)=G(h_z)$ are differentiable at $z_0=\eta(x_0)$ and that $\eta$ is differentiable at $x_0$.
Then
$
    r_X(x_0)
    \leq
    \min
    \left\{
        r_{\mathcal M}(z_0),
        \operatorname{rank}(D\eta_{x_0})
    \right\},
$
and
$
    r_{G,X}(x_0)
    \leq
    \min
    \left\{
        r_{G,\mathcal M}(z_0),
        \operatorname{rank}(D\eta_{x_0})
    \right\}.
$
If the first-order compatibility condition \eqref{eq:functional_first_order_compatibility} holds at $z_0$, then $r_{G,X}(x_0) \leq r_X(x_0).$
\end{proposition}

The proof is given in Appendix~\ref{proof:covariate_induced_rank}.
The proposition separates two distinct sources of dimensional restriction. 
The deformation family determines which hazard directions are available in principle, whereas $D\eta_{x_0}$ determines which combinations of those directions can be generated locally by variation in the covariates.
Local completeness gives a particularly simple consequence.

\begin{corollary}[Completeness under covariate induction]
\label{cor:covariate_induced_completeness}
Under the conditions of Proposition~\ref{prop:covariate_induced_rank_bound}, if $G$ is locally complete for $\mathcal M$ at $z_0=\eta(x_0)$, then $r_{G,X}(x_0) = r_X(x_0)$.
\end{corollary}

Thus, a representation that retains all model-supported first-order hazard directions also retains all directions actually activated by the covariates.
The converse need not hold: a representation may be complete for the covariate-induced variation even if it does not retain every direction available in the larger deformation family.

These ranks describe local structural dimension rather than the magnitude of observed variation.
For example, a covariate map may have rank two even when one direction is activated only weakly, so that the resulting fitted hazard or functional profiles appear nearly one-dimensional over the observed covariate range.
Rank and empirical spread should therefore be interpreted separately.

\begin{remark}[Discrete covariates]
\label{rem:discrete_covariate_rank}
The derivative-based covariate ranks above are local notions for continuously varying covariates.
For mixed covariates, derivatives may be taken with respect to the continuous components conditional on a fixed discrete profile.
Effects of binary or categorical covariates are more naturally represented by finite contrasts between the corresponding fitted hazards or functional representations.
We do not assign a derivative-based covariate rank to purely discrete changes unless a scientifically meaningful smooth embedding is available.
\end{remark}

The resulting hierarchy distinguishes the intrinsic local dimension of the deformation family, the dimension activated by the covariate map, and the dimension visible through a chosen functional representation.
This distinction will be useful for the regression constructions considered
in Section~\ref{sec:power_warp}.

\section{Power-Warp Deformation and Temporal Expressiveness}
\label{sec:power_warp}

Section~\ref{sec:geometry} shows that the independent variation available in a hazard-functional representation is constrained by the temporal deformation directions supported by the underlying survival model.
In particular, AFT models provide only one temporal direction, so multiple timing- and shape-related summaries remain structurally linked through a common scaling coordinate.
This suggests a model-design principle: to enlarge temporal representational capacity, an extension should introduce an additional non-level deformation direction rather than merely add parameters or vertical hazard variation.

We illustrate this principle by extending AFT time scaling to a positive-affine transformation on the log-time scale.
The resulting power-warp mechanism adds one temporal deformation coordinate while retaining the AFT timing component, and is subsequently combined with a log-logistic reference hazard to obtain a parsimonious likelihood-based working specification.

\subsection{From AFT Scaling to Positive-Affine Log-Time Deformation}
\label{subsec:aft_to_power_warp}

An AFT model transforms time through $\phi_\lambda(t)=t e^{-\lambda}$.
On the log-time scale, with $s=\log t$, $\log \phi_\lambda(e^s)=s-\lambda$, so AFT corresponds to a one-dimensional translation family.
It can shift a reference hazard earlier or later but provides only one temporal deformation direction.

To enlarge this temporal capacity while retaining the AFT structure, we consider an extension that remains monotone, contains AFT as a nested special case, and adds only one deformation coordinate.
These requirements do not determine a unique extension.
Here we restrict attention to the positive-affine log-time class,
$
    s-\lambda
    \longrightarrow
    e^\zeta(s-\lambda),
$
$    
    \zeta\in\mathbb R,
$
where $e^\zeta>0$ ensures monotonicity.
Mapping back to the original time scale gives
\begin{equation}
\label{eq:pw}
    \phi_{\lambda,\zeta}(t)
    =
    \exp\left\{
        e^\zeta(\log t-\lambda)
    \right\}
    =
    \left(
        \frac{t}{e^\lambda}
    \right)^{e^\zeta}.
\end{equation}
When $\zeta=0$, \eqref{eq:pw} reduces to the AFT warp
$\phi_{\lambda,0}(t)=t e^{-\lambda}$.
We refer to \eqref{eq:pw} as the \emph{power-warp transformation}.
Within the positive-affine log-time class, this is a minimal extension in the sense that it releases only the fixed unit-slope restriction of AFT while preserving the original timing coordinate $\lambda$.

Let $h_0$ be a positive differentiable reference hazard.
The power-warp transformation induces the hazard family
\begin{equation}
\label{eq:pw-orbit}
    h_{\lambda,\zeta}(t)
    =
    h_0\{\phi_{\lambda,\zeta}(t)\}
    \frac{\partial}{\partial t}
    \phi_{\lambda,\zeta}(t).
\end{equation}
Writing $\ell_0=\log h_0$, the corresponding log-hazard tangent directions at the identity deformation state $(\lambda,\zeta)=(0,0)$ are
$
    \dot\ell_\lambda(t)
    =
    -1-t\ell_0'(t),
$   
and
$
    \dot\ell_\zeta(t)
    =
    1+\log t+t\log t\,\ell_0'(t).
$
The first is the AFT translation direction, whereas the second is the additional log-time dilation direction introduced by power warping.

\begin{proposition}[Local ranks of the power-warp orbit]
\label{prop:pw_rank}
Consider the power-warp hazard family \eqref{eq:pw-orbit}.
Suppose that $h_0$ is positive and continuously differentiable on $[\varepsilon,T]$ and that the projected log-hazard tangent directions $\pi(\dot\ell_\lambda)$ and $\pi(\dot\ell_\zeta)$ are linearly independent in the quotient log-hazard space $\mathcal A_T/V$.
Then
$
    r_{\mathcal M_{\mathrm{PW}}}(0,0)
    =
    r_{\mathrm{temp},\mathcal M_{\mathrm{PW}}}(0,0)
    =
    2.
$
\end{proposition}

\begin{figure}[t]
    \centering
    \includegraphics[width=\textwidth]{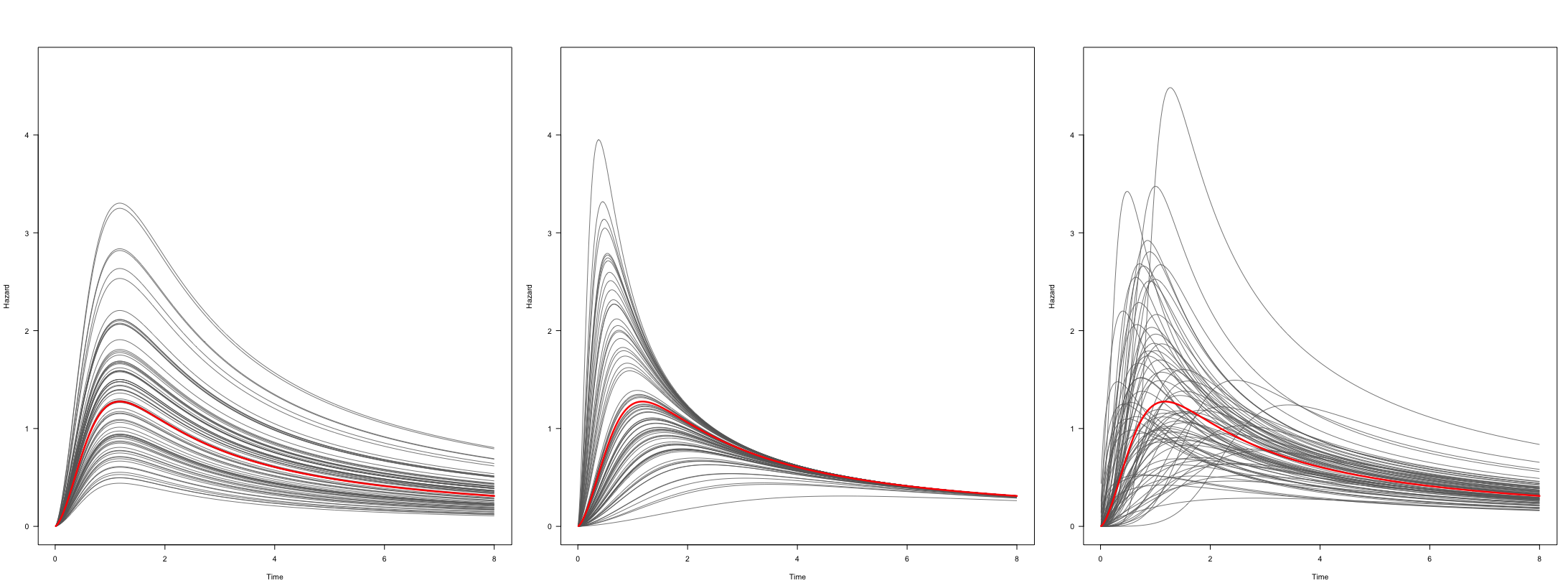}
    \caption{
    Illustration of model-induced hazard variation under proportional-hazards, AFT, and power-warp transformations.
    The red curve denotes a common reference hazard, and the grey curves show representative transformed hazards.
    Proportional-hazards transformations generate only vertical rescaling, whereas AFT transformations additionally permit temporal rescaling through a single temporal deformation coordinate.
    Power-warp transformations introduce an additional log-time dilation coordinate and therefore permit richer variation in hazard timing and shape.
    The displayed trajectories illustrate the transformation mechanisms rather than empirical estimates of local rank.
    }
    \label{fig:transformation_geometry}
\end{figure}

The proof is given in Appendix~\ref{app:powerwarp_rank}.
Together with the AFT rank calculation in Section~\ref{sec:geometry}, Proposition~\ref{prop:pw_rank} shows that power warping adds a genuine non-level temporal deformation direction.
Figure~\ref{fig:transformation_geometry} illustrates how this distinction appears at the level of model-induced hazard trajectories.
Consequently, hazard functionals that are restricted to one local temporal direction under AFT may vary along two directions under the power-warp family, provided that the chosen representation is sensitive to both.

For regression modelling, we use the covariate-dependent power-warp deformation
\begin{equation}
\label{eq:powerwarp_deformation}
    \phi(t;x)
    =
    \left\{
        \frac{t}{\mu(x)}
    \right\}^{\delta(x)},
    \qquad
    \mu(x)>0,
    \quad
    \delta(x)>0.
\end{equation}
Equivalently, with
$
    \lambda(x)=\log\mu(x),
$
$    
    \zeta(x)=\log\delta(x),
$
the corresponding log-time transformation is
$
    \log\phi(e^s;x)
    =
    e^{\zeta(x)}
    \{s-\lambda(x)\}.
$
For any positive differentiable reference hazard $h_0$, this deformation induces
\begin{equation}
\label{eq:powerwarp_hazard_orbit}
    h(t\mid x)
    =
    h_0\left[
        \left\{
            \frac{t}{\mu(x)}
        \right\}^{\delta(x)}
    \right]
    \frac{
        \delta(x)t^{\delta(x)-1}
    }{
        \mu(x)^{\delta(x)}
    }.
\end{equation}
The component $\mu(x)$ controls temporal location, whereas $\delta(x)$ controls log-time dilation.
When $\delta(x)\equiv1$, the construction reduces to the AFT warp $\phi(t;x)=t/\mu(x)$.
The next subsection combines this deformation with a log-logistic reference hazard to obtain the likelihood-based working specification.

\subsection{The Log-Logistic Power-Warp Hazard Specification}
\label{subsec:llpw_specification}

We obtain a tractable likelihood-based specification by combining the power-warp deformation with a unit-scale log-logistic reference distribution,
\begin{equation}
\label{eq:ll_reference_survival}
    S_0(u)
    =
    \frac{1}{1+u^\alpha},
    \qquad
    \alpha>0,
\end{equation}
with reference hazard $h_0(u) = \alpha u^{\alpha-1}/(1+u^\alpha)$.
Under the unit-scale convention, $\alpha$ determines the reference hazard shape, while temporal scale is introduced through the deformation mechanism.
We parameterize the two power-warp components as
\begin{equation}
\label{eq:scaling_component}
    \mu(x)
    =
    \exp\left(\beta_0+x^\top\beta\right),
\end{equation}
and
\begin{equation}
\label{eq:deformation_component}
    \delta(x)
    =
    \exp\left(x^\top\gamma\right),
\end{equation}
where $x$ excludes an intercept.
The component $\mu(x)$ controls temporal location, whereas $\delta(x)$ controls covariate-dependent log-time deformation.
No intercept is included in $\delta(x)$ because, under the log-logistic reference, the effective shape is $\alpha\delta(x)$; an additional deformation intercept would therefore be confounded with $\alpha$.

Combining \eqref{eq:ll_reference_survival} with the power-warp transformation gives
\begin{equation}
\label{eq:llpw_survival}
    S(t\mid x)
    =
    \left[
        1+
        \left\{
            \frac{t}{\mu(x)}
        \right\}^{\alpha\delta(x)}
    \right]^{-1},
\end{equation}
with corresponding hazard
\begin{equation}
\label{eq:llpw_hazard}
    h(t\mid x)
    =
    \frac{
        \alpha\delta(x)t^{\alpha\delta(x)-1}
    }{
        \mu(x)^{\alpha\delta(x)}
        \left[
            1+
            \left\{
                t/\mu(x)
            \right\}^{\alpha\delta(x)}
        \right]
    }.
\end{equation}

We refer to this working specification as the log-logistic power-warp model (LLP).
The parameter $\alpha$ determines the reference shape, $\mu(x)$ the temporal scale, and $\delta(x)$ the covariate-dependent temporal deformation.
The classical log-logistic AFT model is recovered when $\gamma=0$, equivalently $\delta(x)\equiv1$.

\begin{remark}[Relation to flexible parametric survival models]
\label{rem:llp_flexible_parametric}
After reparameterization, the LLP specification is contained in the broader multiparameter survival framework of \citet{burke2020flexible}.
Here, however, the power component arises from a geometry-guided extension of AFT translation by introducing a single additional log-time dilation direction, with the aim of releasing one specific temporal deformation mode rather than pursuing general model flexibility.
Thus, LLP is used as a parsimonious realization of the proposed deformation principle rather than as a new distribution family.
\end{remark}

\begin{remark}[Covariate-induced rank in the LLP specification]
\label{rem:llp_covariate_rank}

The LLP regression provides a direct example of the covariate-induced rank introduced in Section~\ref{sec:geometry}.
Under \eqref{eq:scaling_component}--\eqref{eq:deformation_component}, $D\eta_x = (\beta, \gamma)^{\top}$, so $r_X(x)\leq \operatorname{rank}(D\eta_x)\leq2$.
In particular, if $\gamma=c\beta$ for some constant $c$, covariate variation can activate at most one local hazard direction even though the power-warp mechanism itself supports two.
Thus, model capacity and covariate-activated variation need not coincide.

\end{remark}

The two-dimensional deformation structure also admits a direct interpretation through hazard functionals.
Over the unimodal regime, peak timing and peak height retain the full deformation geometry.

\begin{proposition}[Peak-functional completeness under the LLP specification]
\label{prop:llp_peak_completeness}
Fix the common reference shape parameter $\alpha>0$ and consider the log-logistic power-warp hazard indexed by $(\mu,\delta)$ on $\mathcal D_\alpha = \{(\mu,\delta):\mu>0,\ \alpha\delta>1\}$.
Then the hazard has a unique interior peak, and the map $(\mu,\delta) \longmapsto \bigl(t_{\mathrm{peak}},h_{\mathrm{peak}}\bigr)$ is smoothly invertible on $\mathcal D_\alpha$.
Consequently,
$
    G_{\mathrm{peak}}(h)
    =
    \bigl(
        t_{\mathrm{peak}}(h),
        h_{\mathrm{peak}}(h)
    \bigr)^\top
$
has local functional rank two and is locally complete for the fixed-$\alpha$ LLP hazard family throughout $\mathcal D_\alpha$.
\end{proposition}

The proof is given in Appendix~\ref{proof:llp_peak_completeness}.
In particular, writing $k=\alpha\delta$, the peak coordinates satisfy $t_{\mathrm{peak}}h_{\mathrm{peak}} = \alpha\delta-1$.
Thus the two interpretable functionals provide alternative coordinates for the same two-dimensional deformation represented by $(\mu,\delta)$.

\begin{remark}[Peak constraint under AFT scaling]
\label{rem:aft_peak_constraint}

Under the AFT restriction $\delta\equiv1$, $t_{\mathrm{peak}}h_{\mathrm{peak}} = \alpha-1$.
Hence, for fixed $\alpha>1$, peak timing and peak height are confined to the one-dimensional relation $h_{\mathrm{peak}}=(\alpha-1)/t_{\mathrm{peak}}$.
Although both quantities vary under AFT, they therefore reflect the same temporal scaling coordinate.
The power-warp deformation releases this constraint by allowing $\alpha\delta$ to vary.

\end{remark}

For fixed $\alpha$, Proposition~\ref{prop:llp_peak_completeness} further implies that the entire LLP hazard trajectory is determined by $(t_{\mathrm{peak}},h_{\mathrm{peak}})$.
Additional hazard functionals may provide scientifically distinct descriptions of persistence, cumulative burden, or late-risk allocation, but they do not increase the intrinsic dimension of the model-supported representation.

\subsection{Likelihood Estimation and Inference}
\label{subsec:mle}

Let $(T_i,\Delta_i,x_i)$, $i=1,\ldots,n$, denote independent observations, where
$T_i=\min(T_i^\ast,C_i)$ and
$\Delta_i=\mathbb I(T_i^\ast\le C_i)$.
We assume conditional independent censoring, $T_i^\ast \perp C_i \mid x_i$.
Under the log-logistic power-warp specification, the likelihood is
\[
    L_n(\theta)
    =
    \prod_{i=1}^n
    h(T_i\mid x_i;\theta)^{\Delta_i}
    S(T_i\mid x_i;\theta),
\]
with
$
    \theta
    =
    (\alpha,\beta_0,\beta^\top,\gamma^\top)^\top
    \in
    (0,\infty)\times\mathbb R^{2p+1}.
$
The maximum likelihood estimator is
$
    \widehat\theta_n
    =
    \arg\max_{\theta} \log L_n(\theta).
$
The closed-form log-likelihood, computational parameterization, score vector, and observed information matrix are given in Supplementary Section~S1.

\begin{proposition}[Identifiability]
\label{prop:identifiability}
Consider the log-logistic power-warp model defined by
\eqref{eq:llpw_hazard},
\eqref{eq:scaling_component},
and
\eqref{eq:deformation_component}.
Let $\mathcal X\subseteq\mathbb R^p$ denote the covariate support and assume that $\operatorname{aff}(\mathcal X)=\mathbb R^p$.
Then the model is identifiable, i.e., if
$
    h(t\mid x;\theta) = h(t\mid x;\theta^\star)
$
for all $t>0$ and $x\in\mathcal X$, then $\theta=\theta^\star$.
\end{proposition}

The proof is given in Appendix~\ref{app:identifiability_proof}.
The no-intercept specification for $\delta(x)$ is essential for this result, since an additional deformation intercept would be confounded with the reference shape parameter $\alpha$.
Under correct specification and standard regularity conditions for finite-dimensional maximum likelihood estimation,
$
    \widehat\theta_n
    \xrightarrow{p}
    \theta_0,
$
and
$
    \sqrt n
    (\widehat\theta_n-\theta_0)
    \xrightarrow{d}
    N\{0,I(\theta_0)^{-1}\}.
$
Sufficient regularity conditions and the corresponding large-sample argument are given in Supplementary Section~S2.
Standard errors are obtained from the observed information matrix.

The classical log-logistic AFT model is nested within the LLP specification through $H_0:\gamma=0$, under which $\delta(x)\equiv1$.
The deformation component can therefore be assessed using the likelihood-ratio statistic
\begin{equation}
\label{eq:LRT}
    \Lambda
    =
    2\left\{
        \ell_n(\widehat\theta_{\mathrm{PW}})
        -
        \ell_n(\widehat\theta_{\mathrm{AFT}})
    \right\}.
\end{equation}
Under $H_0$ and the regularity conditions above, $\Lambda \xrightarrow{d} \chi^2_{p_\gamma}$, $p_\gamma=\dim(\gamma)$.


\section{Statistical Recovery of Hazard Functionals}
\label{sec:functional_recovery}

We now turn from model-side representational capacity to the statistical recovery of hazard functionals from fitted hazard trajectories.
The regularity requirements for recovery are related to, but distinct from, the first-order regularity used for the rank analysis in Section~\ref{subsec:functional_rank_bounds}.
For plug-in consistency, the basic requirement is continuity of the functional representation under a topology in which the estimated hazard converges.
For distributional inference, differentiability along the fitted model family becomes relevant, linking the recovery theory directly to the model-path regularity introduced in Section~\ref{subsec:functional_rank_bounds}.
We first establish the general recovery results and then verify uniform hazard convergence for the LLP specification under the likelihood conditions of Section~\ref{subsec:mle}.

\subsection{Consistency and Inference for Hazard-Functional Representations}
\label{subsec:regular_functionals}

Let $\mathbb T=[\varepsilon,\tau]$, with $0<\varepsilon<\tau<\infty$, and let $\mathcal H_{\mathbb T}\subseteq C(\mathbb T)$ denote a class of continuous hazard functions.
For statistical recovery, we equip $\mathcal H_{\mathbb T}$ with the supremum norm
\[
    \|h_1-h_2\|_{\infty,\mathbb T}
    =
    \sup_{t\in\mathbb T}
    |h_1(t)-h_2(t)|.
\]
This topology is distinct from the $L^2$ structure used in Section~\ref{subsec:local_temporal_expressiveness} for tangent-space geometry.

\begin{proposition}[Plug-in consistency of hazard-functional representations]
\label{prop:functional_recovery}
Let $G:\mathcal H_{\mathbb T}\to\mathbb R^K$ be continuous at $h_0$ under the supremum norm.
If
$
    \|\widehat h_n-h_0\|_{\infty,\mathbb T}
    \xrightarrow{p}
    0,
$
then
$
    G(\widehat h_n)
    \xrightarrow{p}
    G(h_0).
$
\end{proposition}

The result follows directly from the continuous mapping theorem.
Uniform convergence is used here as a common sufficient condition rather than a necessary one.
For example, integral functionals may require only suitable $L^1$ convergence, whereas fixed-time hazard evaluations require only pointwise convergence.

The continuity requirement in Proposition~\ref{prop:functional_recovery} is functional specific and concerns consistency rather than first-order differentiability.
Integral functionals, fixed-time hazard evaluations, and the maximum hazard are continuous under the supremum norm, while ratios of continuous integral functionals remain continuous whenever the denominator is non-zero at the target hazard.
Geometry-specific descriptors require additional local stability.
For example, peak timing is consistent under uniform convergence when the target hazard has a unique well-separated maximizer, and threshold-crossing descriptors are stable under isolated transversal crossings.
These conditions are weaker than those required for ordinary first-order differentiation.
In particular, the nondegenerate-peak condition in Proposition~\ref{prop:peak_functional_regularity} is needed for model-path differentiability of peak timing, but not merely for consistency of the estimated maximizer.
Discrete shape descriptors such as modality are generally not continuous under the supremum norm, while derivative-based summaries require stronger convergence controlling the relevant derivatives.

For distributional inference, we return to model-path regularity.
For fixed $x$, the parameter $\theta$ indexes the local hazard family $\theta\mapsto h(\cdot\mid x;\theta)$, so differentiability of
$
    \theta
    \longmapsto
    G\{h(\cdot\mid x;\theta)\}
$
is precisely the corresponding finite-dimensional model-path condition.
Suppose that
\[
    \widehat h_n(\cdot\mid x)
    =
    h(\cdot\mid x;\widehat\theta_n),
\]
where
$
    \sqrt n(\widehat\theta_n-\theta_0)
    \xrightarrow{d}
    N(0,\Sigma_\theta).
$
Define
$
    \Psi_G(\theta;x)
    =
    G\{h(\cdot\mid x;\theta)\}.
$

\begin{corollary}[Asymptotic normality of smooth hazard-functional representations]
\label{coro:ANofGeometry}
If $\Psi_G(\theta;x)$ is differentiable at $\theta_0$, with Jacobian
\[
    \dot\Psi_G(\theta_0;x)
    =
    \left.
    \frac{\partial\Psi_G(\theta;x)}
    {\partial\theta^\top}
    \right|_{\theta=\theta_0},
\]
then
\[
    \sqrt n
    \left[
        G\{\widehat h_n(\cdot\mid x)\}
        -
        G\{h(\cdot\mid x;\theta_0)\}
    \right]
    \xrightarrow{d}
    N_K
    \left(
        0,
        \dot\Psi_G(\theta_0;x)
        \Sigma_\theta
        \dot\Psi_G(\theta_0;x)^\top
    \right).
\]
\end{corollary}

The result follows directly from the multivariate delta method.
For the LLP specification, $\Sigma_\theta=I(\theta_0)^{-1}$ under the likelihood regularity conditions of Section~\ref{subsec:mle}, and the asymptotic covariance can be estimated by
$
    \widehat\Sigma_G(x)
    =
    \dot\Psi_G(\widehat\theta_n;x)
    \widehat\Sigma_\theta
    \dot\Psi_G(\widehat\theta_n;x)^\top.
$
The asymptotic covariance also links statistical uncertainty on the functional scale to the local functional rank introduced in Section~\ref{sec:geometry}.

\begin{corollary}[Rank of the asymptotic functional covariance]
\label{cor:functional_covariance_rank}
Under the conditions of Corollary~\ref{coro:ANofGeometry}, suppose that $\Sigma_\theta$ is positive definite.
Let
$
    V_G(x)
    =
    \dot\Psi_G(\theta_0;x)
    \Sigma_\theta
    \dot\Psi_G(\theta_0;x)^\top.
$
Then
$
    \operatorname{rank}\{V_G(x)\}
    =
    \operatorname{rank}\{\dot\Psi_G(\theta_0;x)\}.
$
Equivalently, when $\theta\mapsto h(\cdot\mid x;\theta)$ is viewed as the local model-induced hazard family of Section~\ref{sec:geometry},
$
    \operatorname{rank}\{V_G(x)\}
    =
    r_{G,\mathcal M}(\theta_0).
$
\end{corollary}

Thus, when parameter uncertainty is nondegenerate, the local functional rank is reflected exactly in the rank of the asymptotic functional covariance.
A singular $V_G(x)$ therefore need not indicate functional rank loss: if the intrinsic model rank is smaller than $K$, a locally complete representation may itself have a rank-deficient asymptotic covariance matrix.

The differentiability requirement in Corollary~\ref{coro:ANofGeometry} is a model-path condition rather than a requirement that $G$ be Fr\'echet differentiable under arbitrary perturbations of the hazard.
Consequently, geometry-specific functionals covered by the local regularity analysis in Section~\ref{subsec:functional_rank_bounds} may also admit ordinary delta-method inference.
In particular, Proposition~\ref{prop:peak_functional_regularity} establishes the required model-path differentiability for peak timing and peak height under a smooth model family with a unique nondegenerate interior maximum.
Nonregular configurations, including tied or flat peaks and tangential threshold crossings, remain outside this ordinary first-order theory.

\subsection{Uniform Hazard Recovery under the Log-Logistic Power-Warp Model}
\label{subsec:llp_uniform_recovery}

Proposition~\ref{prop:functional_recovery} reduces consistent recovery of
hazard-functional representations to uniform recovery of the underlying hazard.
We now verify this condition for the log-logistic power-warp specification.
Let $\widehat h_n(t\mid x) = h(t\mid x;\widehat\theta_n)$, where $\widehat\theta_n$ is the maximum likelihood estimator and $\theta_0$
denotes the true parameter value.
Let $\mathbb T=[\varepsilon,\tau]$, $0<\varepsilon<\tau<\infty$, and let $\mathcal X$ be a compact covariate region.
The restriction $\varepsilon>0$ avoids possible divergence of the LLP hazard near the origin when the effective shape parameter is smaller than one.

\begin{proposition}[Uniform hazard recovery under the LLP model]
\label{prop:llp_uniform_hazard_recovery}
Suppose that
$
    \widehat\theta_n\xrightarrow{p}\theta_0,
$
and that $\theta_0$ lies in the interior of the LLP parameter space.
For any compact covariate region $\mathcal X$ and compact time interval
$\mathbb T=[\varepsilon,\tau]$ with $\varepsilon>0$,
\[
    \sup_{x\in\mathcal X}
    \sup_{t\in\mathbb T}
    \left|
        h(t\mid x;\widehat\theta_n)
        -
        h(t\mid x;\theta_0)
    \right|
    \xrightarrow{p}0.
\]
\end{proposition}

The proof is given in Appendix~\ref{proof:llp_uniform_hazard_recovery}.
The result follows from parameter consistency and joint continuity of the LLP hazard map $(t,x,\theta)\mapsto h(t\mid x;\theta)$ on compact subsets bounded away from $t=0$.
Combining Proposition~\ref{prop:llp_uniform_hazard_recovery} with Proposition~\ref{prop:functional_recovery}, for any fixed $x\in\mathcal X$ and any hazard-functional representation $G:\mathcal H_{\mathbb T}\to\mathbb R^K$ that is continuous at $h(\cdot\mid x;\theta_0)$ under the supremum norm,
\[
    G\{\widehat h_n(\cdot\mid x)\}
    \xrightarrow{p}
    G\{h(\cdot\mid x;\theta_0)\}.
\]
Thus, the LLP likelihood estimator provides the uniform hazard recovery required
for consistent plug-in estimation of regular hazard functionals.

\section{Simulation Studies}
\label{sec:numerical_studies}

This section examines how restrictions on temporal deformation affect finite-sample recovery of hazard trajectories and their functional representations.
The simulations are conducted within the nested log-logistic power-warp family, keeping the reference hazard structure fixed while varying the degree of covariate-dependent departure from the AFT restriction.
This design isolates the consequences of restricting model-supported temporal variation rather than comparing unrelated survival model classes.

We compare the restricted AFT and full power-warp fits under the AFT null and under increasing departures from it.
The simulations address three questions.
First, when the fitted model accommodates the temporal deformation present in the data-generating mechanism, how does hazard and functional recovery improve with sample size?
Second, when an additional temporal deformation direction is present, how does restricting the fit to AFT propagate to recovery error in the hazard trajectory and its functionals?
Third, when the AFT restriction is sufficient, what finite-sample cost is incurred by estimating an unnecessary deformation component?
Likelihood-ratio inference for the AFT restriction is considered as a secondary assessment of whether departures from the restricted deformation structure are detectable from the observed data.

\subsection{Monte Carlo Design}
\label{subsec:monte_carlo_design}

\paragraph{Data-generating mechanisms and fitted hazard families.}

Data were generated from the LLP specification in Section~\ref{subsec:llpw_specification} with $\alpha=2.5$.
The covariates were independently generated as $X_1\sim\mathrm{Unif}(-1,1)$ and $X_2\sim\mathrm{Bernoulli}(0.5)$, with $\beta_0=1$ and $\beta=(0.5,-0.4)^\top$.
Three deformation settings were considered:
\[
\begin{array}{c|c|c}
\text{Scenario} & \gamma & \text{Temporal deformation}\\
\hline
0 & (0,0)^\top & \text{AFT restriction}\\
1 & (0.35,-0.25)^\top & \text{moderate departure from AFT}\\
2 & (0.70,-0.50)^\top & \text{strong departure from AFT}.
\end{array}
\]
Scenario~0 contains no covariate-dependent power deformation, whereas Scenarios~1 and~2 introduce departures from AFT of increasing magnitude.
Independent exponential censoring was calibrated within each scenario to give approximately 30\% censoring.

For each generated dataset, we fitted both the restricted log-logistic AFT submodel, obtained by imposing $\gamma=0$, and the full LLP specification in which $\gamma$ was estimated.
Because the two fitted models share the same log-logistic reference structure, this nested design isolates the recovery consequences of imposing the AFT temporal restriction when additional power-warp deformation is present.

\paragraph{Evaluation design.}

We considered sample sizes $n=300,600,1200$, with $R=1000$ Monte Carlo replications for each design point.
A common independent evaluation sample of $m=500$ covariate profiles was generated once from the covariate distribution and used throughout the replications.
Hazard trajectories and hazard functionals were evaluated at these common profiles, thereby separating model-estimation variability from variation in the evaluation covariates.

Hazard recovery was assessed on $[\varepsilon,T_0]=[0.01,8]$ using integrated absolute error,
\[
\mathrm{IAE}_h(x)
    =
\int_{\varepsilon}^{T_0}
\left|
\widehat h(t\mid x)-h(t\mid x)
\right|\,dt,
\]
and relative integrated squared error.
The corresponding measures were averaged over evaluation profiles and Monte Carlo replications.
Functional recovery was assessed for the fixed-horizon event risks $R_2$ and $R_4$, together with $t_{\mathrm{peak}}$, $h_{\mathrm{peak}}$, $w_{0.5}$, and $\log q_{\mathrm{tail},2,4}$.
These functionals span fixed-horizon risk and several aspects of temporal hazard organization, including timing, intensity, persistence, and relative late-risk allocation.
Definitions are given in Appendix~\ref{app:list_hazard_functionals}.
Functional recovery was summarized by root mean squared error over evaluation profiles and Monte Carlo replications.
To quantify the recovery effect of imposing the AFT restriction, we report the ratio of the RMSE under the full LLP fit to that under the restricted AFT fit, so that ratios below one favor the full specification.
Finally, the likelihood-ratio statistic in \eqref{eq:LRT} was used to test $H_0:\gamma=0$ against $H_1:\gamma\neq0$.
Scenario~0 assesses empirical size under the AFT restriction, whereas Scenarios~1 and~2 assess detection of increasingly strong departures from that restriction.

\subsection{Monte Carlo Results}
\label{subsec:monte_carlo_results}

The observed censoring proportion was close to 30\% throughout.
We first consider recovery of the underlying hazard trajectories.
Table~\ref{tab:mc_hazard_recovery} reports mean integrated absolute error (MIAE) and mean relative integrated squared error (MrISE).

When the AFT restriction is correct, recovery errors decrease with sample size under both fitted models, with smaller finite-sample error under the restricted fit because the additional deformation component is unnecessary.
When covariates activate moderate or strong non-AFT deformation, recovery under the full LLP fit improves steadily with sample size, whereas error under the AFT restriction decreases little and remains substantial.
For example, under strong deformation, MIAE for the LLP fit decreases from 0.293 at $n=300$ to 0.141 at $n=1200$, while the corresponding AFT error remains close to 1.1.
The contrast is even stronger for MrISE.

\begin{table}[!t]
\centering
\small
\caption{
Hazard-trajectory recovery under the restricted AFT and full LLP fits.
MIAE denotes mean integrated absolute error and MrISE denotes mean relative integrated squared error.
The reported ratio is the recovery error under the full LLP fit divided by that under the AFT restriction.
}
\label{tab:mc_hazard_recovery}
\begin{tabular}{llrrrrrr}
\toprule
Scenario & $n$
& \multicolumn{3}{c}{MIAE}
& \multicolumn{3}{c}{MrISE} \\
\cmidrule(lr){3-5}\cmidrule(lr){6-8}
& & AFT restriction & LLP & Ratio
  & AFT restriction & LLP & Ratio \\
\midrule
AFT null    & 300  & 0.257 & 0.369 & 1.43 & 0.0108 & 0.0216 & 2.00 \\
AFT null    & 600  & 0.170 & 0.247 & 1.45 & 0.0047 & 0.0095 & 2.03 \\
AFT null    & 1200 & 0.126 & 0.178 & 1.42 & 0.0026 & 0.0049 & 1.90 \\
\midrule
Moderate PW & 300  & 0.675 & 0.318 & 0.47 & 0.0896 & 0.0219 & 0.24 \\
Moderate PW & 600  & 0.651 & 0.226 & 0.35 & 0.0830 & 0.0109 & 0.13 \\
Moderate PW & 1200 & 0.635 & 0.153 & 0.24 & 0.0788 & 0.0049 & 0.06 \\
\midrule
Strong PW   & 300  & 1.088 & 0.293 & 0.27 & 0.2654 & 0.0230 & 0.09 \\
Strong PW   & 600  & 1.071 & 0.203 & 0.19 & 0.2556 & 0.0108 & 0.04 \\
Strong PW   & 1200 & 1.064 & 0.141 & 0.13 & 0.2510 & 0.0052 & 0.02 \\
\bottomrule
\end{tabular}
\end{table}

We next examine how these model restrictions propagate to the hazard-functional representation.
Table~\ref{tab:mc_functional_recovery} reports the ratio of functional RMSE under the full LLP fit to that under the AFT restriction.
These ratios quantify the recovery consequences of imposing the AFT restriction within the nested design rather than general differences in predictive performance between model classes.

Under the AFT-generating scenario, the ratios are generally above one, reflecting the finite-sample cost of estimating an unnecessary deformation component.
When non-AFT temporal deformation is activated, the full LLP fit yields substantially smaller functional recovery errors, with the contrast increasing with deformation magnitude and sample size.
The largest relative differences occur for several functionals describing temporal organization.
Under strong deformation and $n=1200$, the RMSE ratios are 0.14 for $t_{\mathrm{peak}}$, 0.16 for $w_{0.5}$, and 0.11 for $\log q_{\mathrm{tail},2,4}$, compared with 0.24 and 0.25 for the fixed-horizon risks $R_2$ and $R_4$.
Thus, in this design, restricting covariate-induced temporal variation to AFT scaling has particularly pronounced consequences for recovery of temporal hazard features.

\begin{table}[!t]
\centering
\small
\caption{
Hazard-functional recovery under the nested deformation design.
Entries are RMSE ratios, defined as the RMSE under the full LLP fit divided by that under the AFT restriction.
Ratios below one indicate smaller recovery error under the full LLP specification.
}
\label{tab:mc_functional_recovery}
\begin{tabular}{llrrrrrr}
\toprule
Scenario & $n$
& $t_{\mathrm{peak}}$
& $h_{\mathrm{peak}}$
& $w_{0.5}$
& $\log q_{\mathrm{tail},2,4}$
& $R_2$
& $R_4$ \\
\midrule
AFT null    & 300  & 1.10 & 1.41 & 1.13 & 1.37 & 1.05 & 1.11 \\
AFT null    & 600  & 1.09 & 1.42 & 1.13 & 1.35 & 1.05 & 1.13 \\
AFT null    & 1200 & 1.10 & 1.39 & 1.12 & 1.34 & 1.05 & 1.10 \\
\midrule
Moderate PW & 300  & 0.62 & 0.50 & 0.71 & 0.45 & 0.70 & 0.78 \\
Moderate PW & 600  & 0.47 & 0.37 & 0.56 & 0.32 & 0.58 & 0.63 \\
Moderate PW & 1200 & 0.33 & 0.26 & 0.41 & 0.22 & 0.43 & 0.45 \\
\midrule
Strong PW   & 300  & 0.28 & 0.62 & 0.33 & 0.23 & 0.45 & 0.49 \\
Strong PW   & 600  & 0.21 & 0.41 & 0.23 & 0.16 & 0.34 & 0.36 \\
Strong PW   & 1200 & 0.14 & 0.30 & 0.16 & 0.11 & 0.24 & 0.25 \\
\bottomrule
\end{tabular}
\end{table}

As a secondary assessment, Table~\ref{tab:mc_lrt} reports likelihood-ratio inference for the AFT restriction.
Under the null scenario, empirical rejection rates are close to the nominal 5\% level and the mean likelihood-ratio statistic is close to the $\chi^2_2$ reference mean of 2.
Under moderate and strong deformation, rejection probabilities increase with sample size and deformation strength.
Thus, in these controlled scenarios, the departures producing the recovery differences above are also detectable by standard likelihood-based inference.

\begin{table}[!t]
\centering
\small
\caption{
Likelihood-ratio assessment of the AFT restriction.
Entries are empirical rejection rates at level 0.05, with the mean likelihood-ratio statistic in parentheses.
}
\label{tab:mc_lrt}
\begin{tabular}{lccc}
\toprule
Scenario & $n=300$ & $n=600$ & $n=1200$ \\
\midrule
AFT null      & 0.064 (2.06)  & 0.045 (1.92)   & 0.052 (2.03) \\
Moderate PW   & 0.962 (18.87) & 1.000 (37.21)  & 1.000 (70.00) \\
Strong PW     & 1.000 (68.81) & 1.000 (133.79) & 1.000 (267.41) \\
\bottomrule
\end{tabular}
\end{table}

Overall, the simulations illustrate how restrictions on covariate-activated temporal deformation propagate from the fitted hazard family to recovery of both hazard trajectories and their functional representations.
When the fitted model accommodates the deformation present in the data-generating mechanism, recovery improves with sample size.
When the data contain non-AFT temporal deformation but the fit is restricted to a single AFT scaling direction, substantial hazard and functional recovery errors persist, whereas estimating the additional deformation component incurs a finite-sample cost when the AFT restriction is correct.
These results provide a controlled illustration of the recovery consequences of model-induced temporal restrictions rather than a general comparison of predictive performance across survival model classes.

\section{Application to the Lung Data}
\label{sec:applications}

The preceding sections show that fitted survival models induce both hazard-functional representations and structural constraints on their joint variation.
For an individual with covariate profile $x$, the estimated vector
$
    G\{\widehat h(\cdot\mid x)\}
$
summarizes interpretable features of the fitted hazard trajectory and may be examined individually or jointly as a multivariate representation of temporal risk.
The functionals used below are defined in Appendix~\ref{app:list_hazard_functionals}.

For descriptive interpretation, selected functionals may also be standardized or combined into lower-dimensional profile summaries.
We refer to the resulting collection as an individual risk profile.
These summaries organize scientifically distinct aspects of the fitted hazard trajectory, but they need not represent independent dimensions of model-supported variation.
Because the construction operates on fitted hazard trajectories, it is not specific to the LLP specification and can be applied to other survival models that provide individual hazard estimates on a common time domain.

We illustrate the framework using the \texttt{lung} dataset from the \texttt{survival} package, with particular attention to how model-induced restrictions appear in the resulting hazard-functional representations.
After excluding observations with missing covariates, the analysis included 227 patients, of whom 164 experienced the event.
The covariates were age, sex, and ECOG performance score, and follow-up time was measured in years.
The analysis is intended as a full-data illustration of how model-induced restrictions appear in fitted hazard-functional representations and how fitted mortality differences can be further described through temporal risk features.

We fitted the LLP specification as the working model for individual hazard trajectories together with its nested log-logistic AFT submodel.
Both models converged successfully.
The LLP fit had a higher maximized log-likelihood and lower AIC than the AFT fit (Table~\ref{tab:lung_model_fit}).
The likelihood-ratio statistic for the AFT restriction was 9.10 on 3 degrees of freedom ($p=0.028$), providing some evidence against pure AFT time scaling.

\begin{table}[ht]
\centering
\caption{Model fit summaries for the lung data.}
\label{tab:lung_model_fit}
\begin{tabular}{lrrr}
\hline
Model & Log-likelihood & AIC & $\hat\alpha$ \\
\hline
AFT & $-169.91$ & $349.81$ & $1.87$ \\
LLP & $-165.36$ & $346.71$ & $1.87$ \\
\hline
\end{tabular}
\end{table}

Individual fitted hazards were evaluated on a common time grid extending from one day to the 95th percentile of observed follow-up and converted into the hazard functionals defined in Appendix~\ref{app:list_hazard_functionals}.
Under the LLP fit, the estimated deformation component $\widehat\delta(x)$ had median 1.00, interquartile range 0.89--1.18, and range 0.62--1.69.
Since $\widehat\alpha=1.87$, the corresponding effective shape
$
    \widehat k(x)=\widehat\alpha\,\widehat\delta(x)
$
ranged approximately from 1.16 to 3.16.
Thus, all fitted LLP profiles satisfy the unimodality condition $\widehat k(x)>1$ required by Proposition~\ref{prop:llp_peak_completeness}.
The peak pair
$
    (t_{\mathrm{peak}},h_{\mathrm{peak}})
$
therefore provides complete functional coordinates for the fitted two-dimensional power-warp deformation family.

For the applied representation, fixed-horizon event risk was evaluated at one year, the early-risk window ended at 0.5 years, and the late-risk window began at one year.
For graphical and descriptive interpretation, selected functionals were additionally organized into three standardized profile summaries.
Let $Z_i(g)$ denote the standardized value of functional $g$ across individuals.
We define
$
\mathrm{Timing}_i
=
Z_i(t_{\mathrm{peak}})
$,
$
\mathrm{Intensity}_i
=
Z_i(h_{\mathrm{peak}})
$,
and
$
\mathrm{Persistence}_i
=
0.5
\{
Z_i(w_{0.5})
+
Z_i(\log q_{\mathrm{tail}})
\}.
$
These quantities provide different scientific descriptions of temporal risk, but they should not be interpreted as three independent deformation dimensions.
In particular, within the fitted LLP family, Proposition~\ref{prop:llp_peak_completeness} implies that the complete model-supported deformation state is already determined by peak timing and peak intensity; Persistence remains an interpretable descriptive summary without increasing the intrinsic dimension of the representation.

Table~\ref{tab:lung_functional_summary} summarizes the LLP-based functional representation.
The median estimated one-year event risk was 0.65, while the temporal descriptors showed substantial between-subject variation.
Median peak time was 0.75 years, median half-peak width was 1.77 years, and median tail ratio was 2.84.
These summaries illustrate that fitted hazards can be described through several scientifically distinct temporal features even when those features do not correspond to independent model-supported directions.

\begin{table}[ht]
\centering
\caption{Summary of hazard functionals under the LLP fit.}
\label{tab:lung_functional_summary}
\begin{tabular}{lrrrr}
\hline
Functional & Mean & Median & IQR & Range \\
\hline
One-year risk & 0.58 & 0.65 & 0.47--0.68 & 0.18--0.91 \\
One-year cumulative hazard & 0.95 & 1.05 & 0.64--1.15 & 0.20--2.38 \\
Peak time & 0.83 & 0.75 & 0.51--1.05 & 0.22--2.01 \\
Peak height & 1.28 & 1.23 & 0.88--1.52 & 0.50--3.20 \\
Early hazard mass & 0.37 & 0.37 & 0.19--0.48 & 0.03--1.32 \\
Late hazard mass & 0.98 & 0.94 & 0.78--1.13 & 0.48--1.83 \\
High-risk width & 1.72 & 1.77 & 1.56--1.88 & 0.98--2.01 \\
Tail ratio & 4.09 & 2.84 & 1.88--4.61 & 0.91--25.17 \\
\hline
\end{tabular}
\end{table}

To illustrate how model structure propagates into hazard-functional representation space, we additionally fitted a Cox proportional hazards model using the same covariates.
The Cox model is used as a structural reference rather than as an additional likelihood competitor.
Its baseline cumulative hazard was estimated by the Breslow estimator \citep{lin2007breslow}, and the increments were smoothed by a cubic smoothing spline on the log-hazard scale (\texttt{spar}=0.6) to obtain continuous hazard trajectories for functional extraction.
This smoothing preserves the proportional-hazards restriction that individual hazards differ only through vertical rescaling.

\begin{figure}[!htb]
\centering
\includegraphics[
    width=0.85\textwidth,
    height=0.62\textheight,
    keepaspectratio
]{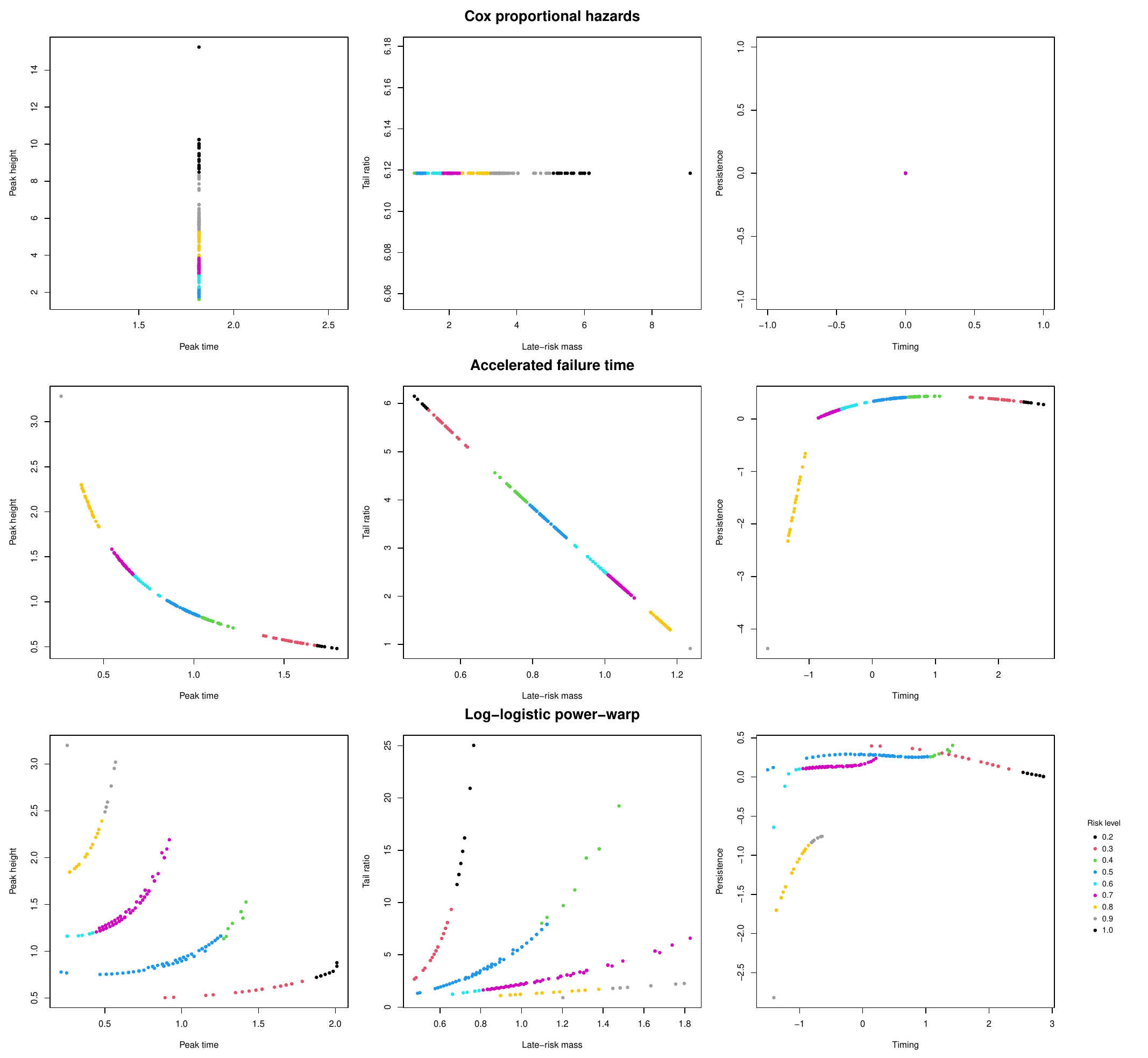}
\caption{
Hazard-functional representations for the lung data under Cox proportional hazards, AFT, and LLP fits.
Rows correspond to Cox, AFT, and LLP models.
Columns display peak timing versus peak intensity, late-risk mass versus tail ratio, and the standardized Timing and Persistence summaries.
Each point represents one patient, with colour indicating fitted one-year event risk rounded to the nearest 0.1.
The first column illustrates particularly direct structural restrictions: proportional hazards fixes peak timing, the AFT model confines peak timing and height to a one-dimensional relation, whereas under the LLP fit the two peak functionals provide complete coordinates for the power-warp deformation state.
The point clouds themselves are not empirical estimates of local rank.
}
\label{fig:lung_hazard_functional_representation}
\end{figure}

Figure~\ref{fig:lung_hazard_functional_representation} compares the representations obtained from the Cox, AFT, and LLP fits for the same 227 covariate profiles.
The first column displays peak timing against peak intensity, the second late-risk mass against tail ratio, and the third the standardized Timing and Persistence summaries.
Each point represents one patient and is coloured by fitted one-year event risk, rounded to the nearest 0.1.

The three rows exhibit the structural restrictions described in Section~\ref{sec:geometry}.
Under proportional hazards, covariate effects change hazard level but not temporal organization, so peak timing is essentially fixed while peak intensity varies.
Under the log-logistic AFT model, temporal variation is available, but peak timing and peak intensity remain governed by the same scaling coordinate.
By Remark~\ref{rem:aft_peak_constraint}, $t_{\mathrm{peak}}h_{\mathrm{peak}} = \alpha-1.$
For the fitted AFT model, $\widehat\alpha=1.87$, so the attainable peak pairs satisfy approximately $t_{\mathrm{peak}}h_{\mathrm{peak}}=0.87.$
The one-dimensional pattern in the first-column AFT panel is therefore a direct consequence of the model restriction rather than merely an empirical association between fitted summaries.

The LLP fit has a different interpretation.
Because Proposition~\ref{prop:llp_peak_completeness} applies to all fitted covariate profiles, the pair $(t_{\mathrm{peak}},h_{\mathrm{peak}})$
is not merely a two-dimensional display of selected summaries: it provides complete functional coordinates for the fitted power-warp deformation state.
Variation in $t_{\mathrm{peak}}h_{\mathrm{peak}} = \alpha\delta-1$ records variation in the effective shape parameter, while variation in peak timing also reflects temporal scale.
Thus the additional deformation mechanism releases the one-dimensional peak constraint imposed by AFT.

The realized point clouds should nevertheless not be interpreted as estimates of intrinsic model rank.
The model determines which local deformation directions are available, whereas the variation actually observed across individuals also depends on the fitted regression coefficients and the empirical covariate distribution.
The figure therefore illustrates model-induced structural constraints and their realization in this dataset rather than estimating the dimension of the underlying model family from the scatter itself.
The colouring by one-year event risk further shows that similar fixed-horizon risks need not uniquely determine the corresponding temporal risk pattern.

The remaining panels emphasize a complementary point.
Functionals such as late-risk mass, tail ratio, and Persistence provide scientifically different descriptions of the fitted hazard trajectory, even though under the two-dimensional LLP family they cannot add a third independent deformation direction once the complete peak coordinates are known.
Structural redundancy in this sense does not make additional functionals uninformative; rather, it distinguishes additional interpretation from additional intrinsic model-supported variation.

We next use these descriptive summaries to characterize how the fitted multivariable survival relationship is expressed through Timing, Intensity, and Persistence.
Each derived summary was linearly projected on standardized age and ECOG together with the sex indicator used in the survival model, with these projections interpreted descriptively rather than as independent second-stage tests.

Higher ECOG performance score was associated with higher fitted one-year mortality risk, expressed mainly through an earlier and more intense hazard peak and relatively less allocation of hazard to later follow-up.
A one-standard-deviation increase in ECOG corresponded to a 0.20-year earlier peak, a 0.44 increase in peak hazard, and a 0.25 decrease in log late-to-early hazard ratio, while high-risk width changed little.

Female patients showed a different pattern, with lower fitted one-year mortality risk, later and lower hazard peaks, and relatively greater allocation of hazard to later follow-up.
After adjustment for age and ECOG, female sex corresponded to a 0.46-year later peak, a 0.46 lower peak hazard, and a 0.65 higher log late-to-early hazard ratio, again with little change in high-risk width.
Median one-year risks were 0.67 and 0.47 for male and female patients, respectively, while median peak times were 0.58 and 1.15 years and median log tail ratios were 0.76 and 1.63.
Together, these patterns illustrate that differences in fitted short-term risk can reflect distinct temporal organizations of the underlying hazard trajectory.

These post-hoc summaries provide a more specific temporal interpretation of the fitted covariate--survival relationship beyond higher or lower event risk.
Because all functionals are derived from the same fitted survival model, scientifically distinct summaries should not be interpreted as independent sources of statistical information.
In the present LLP application, Proposition~\ref{prop:llp_peak_completeness} makes this distinction explicit: peak timing and peak intensity already form a complete two-dimensional coordinate system for the fitted deformation family, while quantities such as Persistence, tail allocation, and fixed-horizon risk provide alternative descriptions of that same model-supported hazard variation.
The post-hoc projections should therefore be viewed as decompositions of the fitted survival relationship rather than as independent inferential evidence; formal downstream inference would additionally require propagation of first-stage estimation uncertainty.

Overall, the lung-data example illustrates three distinct levels of representation.
Model structure determines the intrinsic hazard variation that is available; selected hazard functionals determine how that variation is represented; and additional profile summaries provide scientifically useful descriptions without necessarily increasing representational dimension.
The comparison among Cox, AFT, and LLP makes these distinctions visible in a common functional space, while the LLP fit shows how a pair of interpretable hazard functionals can serve as complete coordinates for model-supported temporal deformation.

\section{Discussion}
\label{sec:discussion}

This paper develops a hazard-functional framework for representing and interpreting temporal variation in fitted survival trajectories.
Rather than relying on a single scalar summary, the framework uses scientifically interpretable functionals of the conditional hazard trajectory and studies the structure of their joint variation.
The central statistical point is that the number of reported functionals need not equal the number of independent hazard directions supported by the underlying survival model.
Model-induced tangent spaces characterize the available first-order hazard variation, while functional rank quantifies how much of that variation remains visible after mapping through $G$.
Under the first-order compatibility condition, the local model rank decomposes exactly into variation retained by the representation and model-supported directions that are locally invisible to it.
This yields a model-relative notion of local completeness, while covariate-induced rank further distinguishes variation available from the model from that actually activated by the regression structure.
The resulting ranks are properties of the local hazard family rather than artifacts of a particular regular parameterization.
Together, these results show that adding more hazard summaries cannot create independent variation that is absent from the underlying model.

The power-warp construction provides a concrete illustration of this model-design principle.
Under the regularity conditions considered here, AFT scaling supports a single temporal deformation direction, whereas positive-affine deformation on the log-time scale can support a second non-level temporal direction.
Combining this mechanism with a log-logistic reference hazard yields the parsimonious LLP specification containing the classical log-logistic AFT model as a nested special case.
For a fixed reference shape over the unimodal regime, peak timing and peak height provide complete coordinates for the corresponding two-dimensional LLP hazard family.
At the same time, the covariate-induced rank result shows that a two-dimensional deformation mechanism need not produce two-dimensional variation across individuals if the covariate map activates only a lower-dimensional subset of those directions.
The LLP specification is therefore intended as a tractable illustration of geometry-guided model extension rather than as a new general-purpose survival distribution.

The recovery results address the separate question of whether model-supported hazard functionals can be estimated reliably from data.
Uniform hazard recovery provides a general sufficient condition for consistent plug-in estimation of continuous functionals, while geometry-specific descriptors require additional local stability.
For smooth parameter-induced representations, delta-method inference propagates parameter uncertainty to the functional scale.
When parameter uncertainty is nondegenerate, the rank of the asymptotic functional covariance coincides with the local functional rank.
Thus, rank deficiency of the functional covariance may reflect intrinsic model-supported dependence among reported functionals rather than failure of an otherwise complete representation.

The framework also provides a basis for interpreting fitted covariate--survival relationships on the functional scale.
Given censored survival observations $\mathcal D_n = \{(Y_i,\Delta_i,X_i)\}_{i=1}^n$, a fitted survival model induces the sequence
$
\mathcal D_n
\longrightarrow
\widehat h_n(\cdot\mid x)
\longrightarrow
\widehat G_n(x)
    =
    G\{\widehat h_n(\cdot\mid x)\}.
$
Hazard-functional extraction can therefore be viewed as a model-based transformation of fitted survival information into interpretable temporal-risk targets.
Whereas conventional analyses often describe fitted covariate differences through regression parameters or risks at selected horizons, the functional representation can further characterize how those differences are expressed through timing, intensity, persistence, or early-versus-late risk allocation.
The lung-data example illustrates both this interpretive role and the different structural restrictions induced by Cox, AFT, and LLP models.
The resulting empirical point clouds should not be interpreted as estimates of local rank, because their spread also depends on the fitted covariate map and the observed covariate distribution.
Likewise, hazard-derived summaries obtained from the same fitted model provide a decomposition of the fitted covariate--survival relationship rather than independent evidence of association.

Several limitations define the scope of the present framework.
First, the rank theory is local and first-order and therefore describes directions available near a given hazard state rather than the full nonlinear geometry of the attainable hazard family.
Second, the current development focuses on transformation-induced hazard families, while the LLP specification deliberately permits only two temporal deformation directions and does not accommodate arbitrary local bending, multimodality, or irregular time-varying effects.
Third, the representation $G$ is scientifically chosen rather than universally determined, and distinct scientifically useful functionals may remain structurally dependent.
Fourth, the simulations examine controlled departures from AFT within a common reference family, and the data analysis is intended to illustrate model-induced functional structure rather than establish general predictive superiority or external validity.
More generally, greater functional representational capacity does not by itself imply improved discrimination, calibration, or decision utility.

A further inferential problem arises when estimated functionals such as $\widehat G_n(X_i)$ are subsequently used as outcomes or targets in downstream population-level analyses.
Such quantities are generated variables, so uncertainty from estimation of the first-stage survival trajectory must generally be propagated into the downstream analysis \citep{murphy2002estimation,wang2020postprediction}.
Additional dependence arises when the same observations contribute to both stages, and regular smooth functionals may require different methods from nonregular geometric quantities such as peak locations or threshold crossings.
For flexible, semiparametric, or machine-learning hazard estimators, valid downstream inference may therefore require influence-function methods, sample splitting, or orthogonalization.
Other extensions include principled selection of hazard functionals according to scientific relevance, statistical stability, and structural redundancy, richer deformation systems, sensitivity to reference-hazard misspecification, and out-of-sample assessment of functional representations.

More broadly, the framework suggests that survival-model flexibility should be assessed not only through parameterization or predictive fit, but also through the hazard variation that a model can support and expose through scientifically meaningful summaries.
Scalar risk measures remain useful for ranking, stratification, and decision-making, but they need not uniquely characterize the temporal organization of risk.
Hazard-functional representations provide an intermediate statistical layer between fitted survival trajectories and scientific interpretation, while model-induced geometry clarifies the independent variation that this layer can retain.

\bibliographystyle{plainnat}
\bibliography{ref} 

\appendix

\section{Proof of Proposition~\ref{prop:overall_temporal_ranks_basic_orbits}}
\label{app:proof_overall_temporal_ranks_basic_orbits}

\begin{proof}
All calculations are made at the identity deformation state. 
Write $\ell_0=\log h_0$. Since $h_0$ is strictly positive on $[\varepsilon,T_0]$, the log-hazard derivatives below are well defined.
For the proportional hazards orbit, differentiating with respect to $\eta$ at $\eta=0$ gives
\[
    \left.
    \frac{\partial h_\eta(t)}{\partial \eta}
    \right|_{\eta=0}
    =
    h_0(t).
\]
Hence $\mathcal T_{\mathrm{PH}}(0) = \operatorname{span}\{h_0\}$, $r_{\mathrm{PH}}(0)=1$.
On the log-hazard scale,
$
    \ell_\eta(t)
    =
    \log h_\eta(t)
    =
    \eta+\ell_0(t),
$
so
\[
    \left.
    \frac{\partial \ell_\eta(t)}{\partial \eta}
    \right|_{\eta=0}
    =
    1.
\]
Since $1\in V=\operatorname{span}\{1\}$, its quotient projection is zero:
$
    \pi(1)=0,
$
and therefore
$
    r_{\mathrm{temp},\mathrm{PH}}(0)=0.
$
For the AFT orbit, differentiating with respect to $\lambda$ at $\lambda=0$ gives
\[
    \left.
    \frac{\partial h_\lambda(t)}{\partial \lambda}
    \right|_{\lambda=0}
    =
    -h_0(t)-t h_0'(t).
\]
Thus
$
    \mathcal T_{\mathrm{AFT}}(0)
    =
    \operatorname{span}\{-h_0-th_0'\}.
$
The nondegeneracy condition that $t\ell_0'(t)$ is not constant implies that
$-h_0-th_0'$ is not identically zero. Hence
$
    r_{\mathrm{AFT}}(0)=1.
$
On the log-hazard scale,
\[
    \ell_\lambda(t)
    =
    \log h_\lambda(t)
    =
    -\lambda+\ell_0(te^{-\lambda}).
\]
Differentiating at $\lambda=0$ gives
\[
    \left.
    \frac{\partial \ell_\lambda(t)}{\partial \lambda}
    \right|_{\lambda=0}
    =
    -1-t\ell_0'(t).
\]
The constant term $-1$ is removed in the quotient space, so
\[
    \pi\{-1-t\ell_0'(t)\}
    =
    \pi\{-t\ell_0'(t)\}.
\]
Since $t\ell_0'(t)$ is not constant on $[\varepsilon, T_0]$, $\pi\{-t\ell_0'(t)\}$ is nonzero in $\mathcal A_T/V$. 
Therefore $r_{\mathrm{temp},\mathrm{AFT}}(0)=1.$
For the AFT-plus-level orbit, differentiating at $(\lambda,\eta)=(0,0)$ gives
\[
    \left.
    \frac{\partial h_{\lambda,\eta}(t)}{\partial \lambda}
    \right|_{(0,0)}
    =
    -h_0(t)-t h_0'(t),
    \qquad
    \left.
    \frac{\partial h_{\lambda,\eta}(t)}{\partial \eta}
    \right|_{(0,0)}
    =
    h_0(t).
\]
Hence
$
    \mathcal T_{\mathrm{AFT+L}}(0,0)
    =
    \operatorname{span}\{-h_0-th_0',h_0\}.
$
To show that these two directions are linearly independent, suppose that
$
    -h_0(t)-t h_0'(t)=c h_0(t)
$
for some constant $c$ and all $t\in[\varepsilon, T_0]$. 
Dividing by $h_0(t)>0$ gives
\[
    -1-t\frac{h_0'(t)}{h_0(t)} = c,
\]
or equivalently $ t\ell_0'(t)=-(1+c)$, which is constant. 
This contradicts the assumed nondegeneracy condition.
Therefore the two directions are linearly independent, and $r_{\mathrm{AFT+L}}(0,0)=2$.

On the log-hazard scale,
\[
    \ell_{\lambda,\eta}(t)
    =
    \eta-\lambda+\ell_0(te^{-\lambda}).
\]
Thus
\[
    \left.
    \frac{\partial \ell_{\lambda,\eta}(t)}{\partial \eta}
    \right|_{(0,0)}
    =
    1,
    \qquad
    \left.
    \frac{\partial \ell_{\lambda,\eta}(t)}{\partial \lambda}
    \right|_{(0,0)}
    =
    -1-t\ell_0'(t).
\]
After projection to $\mathcal A_T/V$, $\pi(1)=0$, $ \pi\{-1-t\ell_0'(t)\} = \pi\{-t\ell_0'(t)\}$.
As above, $\pi\{-t\ell_0'(t)\}$ is nonzero under the nondegeneracy condition.
Hence the projected temporal tangent space is one-dimensional, and $r_{\mathrm{temp},\mathrm{AFT+L}}(0,0)=1.$
The proof of this proposition is complete.
\end{proof}

\section{Proof of Proposition~\ref{prop:peak_functional_regularity}}
\label{proof:peak_functional_regularity}

\begin{proof}
Let
$
    t_p=t_{\mathrm{peak}}(h_{z_0}).
$
By assumption,
$
    h_{z_0}'(t_p)=0,
$
$
    h_{z_0}''(t_p)<0.
$
Define
$
    F(z,t)
    =
    \partial h_z(t)/\partial t.
$
Then
\[
    F(z_0,t_p)=0,
    \qquad
    \frac{\partial F}{\partial t}(z_0,t_p)
    =
    h_{z_0}''(t_p)\neq0.
\]
The implicit function theorem therefore gives a differentiable local map
$z\mapsto t_p(z)$ satisfying
$
    F\{z,t_p(z)\}=0,
$
with $t_p(z_0)=t_p$. By uniqueness and local stability of the maximum, $t_p(z)=t_{\mathrm{peak}}(h_z)$ in a neighbourhood of $z_0$.

For any $a\in\mathbb R^q$, let
$
    v_a(t)=Dh_{z_0}[a](t).
$
Differentiating
$
    h_z'\{t_p(z)\}=0
$
at $z_0$ in the direction $a$ gives
$
    v_a'(t_p)
    +
    h_{z_0}''(t_p)
    D t_{\mathrm{peak}}(z_0)[a]
    =
    0.
$
Hence
$
    D t_{\mathrm{peak}}(z_0)[a]
    = -v_a'(t_p)/h_{z_0}''(t_p).
$
For the peak height,
$
    h_{\mathrm{peak}}(h_z)
    =
    h_z\{t_{\mathrm{peak}}(h_z)\}.
$
Differentiating in the direction $a$ yields
\[
\begin{aligned}
    D h_{\mathrm{peak}}(z_0)[a]
    &=
    v_a(t_p)
    +
    h_{z_0}'(t_p)
    D t_{\mathrm{peak}}(z_0)[a] \\
    &=
    v_a(t_p),
\end{aligned}
\]
since $h_{z_0}'(t_p)=0$.
Therefore the peak-based representation
$
    G_{\mathrm{peak}}(h)
    =
    \{t_{\mathrm{peak}}(h),h_{\mathrm{peak}}(h)\}^{\top}
$
is differentiable along the model-induced family at $z_0$.
Finally, if $ Dh_{z_0}[a]=0, $ then $v_a\equiv0$, so $v_a(t_p)=0$, $v_a'(t_p)=0$.
Consequently,
$
    D t_{\mathrm{peak}}(z_0)[a]=0,
$   
$
    D h_{\mathrm{peak}}(z_0)[a]=0,
$
and hence
\[
    D\Psi_{G_{\mathrm{peak}}}(z_0)[a]=0.
\]
Thus the first-order compatibility condition
\eqref{eq:functional_first_order_compatibility} is satisfied.
\end{proof}

\section{Proof of Theorem~\ref{thm:functional_rank_bound}}
\label{proof:geom_bound}

\begin{proof}
We first prove part~(i).
Write
\[
    A = Dh_{z_0}:\mathbb R^q\to\mathcal H,
    \qquad
    B = D\Psi_G(z_0):\mathbb R^q\to\mathbb R^K.
\]
By definition,
$
    \operatorname{Im}(A)
    =
    \mathcal T_{\mathcal M}(z_0).
$
The first-order compatibility condition states that
\[
    A[a]=0
    \quad\Longrightarrow\quad
    B[a]=0.
\]

For $v\in\mathcal T_{\mathcal M}(z_0)$, choose any
$a\in\mathbb R^q$ such that $v=A[a]$, and define
$
    L_{G,\mathcal M}(z_0)[v]
    =
    B[a].
$
This definition is well defined.
Indeed, if $A[a_1]=A[a_2]$, then
$
    A[a_1-a_2]=0,
$
and the compatibility condition implies
$
    B[a_1-a_2]=0.
$
Hence $B[a_1]=B[a_2]$.
Linearity follows from the linearity of $A$ and $B$.
Moreover, since every element of
$\mathcal T_{\mathcal M}(z_0)$ lies in the image of $A$, this construction is the unique linear map satisfying
$
    L_{G,\mathcal M}(z_0)\{Dh_{z_0}[a]\}
    =
    D\Psi_G(z_0)[a].
$

Because $A$ maps onto $\mathcal T_{\mathcal M}(z_0)$,
$
    \operatorname{Im}\{L_{G,\mathcal M}(z_0)\}
    =
    \operatorname{Im}\{D\Psi_G(z_0)\}.
$
Therefore,
$
    r_{G,\mathcal M}(z_0)
    =
    \operatorname{rank}\{L_{G,\mathcal M}(z_0)\}.
$
Since $\mathcal T_{\mathcal M}(z_0)$ is finite-dimensional, the rank--nullity theorem gives
\[
\begin{aligned}
    r_{G,\mathcal M}(z_0)
    &=
    \dim\mathcal T_{\mathcal M}(z_0)
    -
    \dim
    \left\{
        v\in\mathcal T_{\mathcal M}(z_0):
        L_{G,\mathcal M}(z_0)[v]=0
    \right\} \\
    &=
    r_{\mathcal M}(z_0)
    -
    \dim
    \left\{
        v\in\mathcal T_{\mathcal M}(z_0):
        L_{G,\mathcal M}(z_0)[v]=0
    \right\}.
\end{aligned}
\]
In particular,
$
    r_{G,\mathcal M}(z_0)
    \leq
    r_{\mathcal M}(z_0).
$

If $G$ is additionally Fr\'echet differentiable at $h_{z_0}$, the chain rule
gives
\[
    D\Psi_G(z_0)
    =
    DG_{h_{z_0}}
    \circ
    Dh_{z_0}.
\]
Thus, for any
$v=Dh_{z_0}[a]\in\mathcal T_{\mathcal M}(z_0)$,
\[
\begin{aligned}
    L_{G,\mathcal M}(z_0)[v]
    &=
    D\Psi_G(z_0)[a] \\
    &=
    DG_{h_{z_0}}\{Dh_{z_0}[a]\} \\
    &=
    DG_{h_{z_0}}[v].
\end{aligned}
\]
Hence
$
    L_{G,\mathcal M}(z_0)
    =
    DG_{h_{z_0}}
    \big|_{\mathcal T_{\mathcal M}(z_0)},
$
and consequently
\[
    \left\{
        v\in\mathcal T_{\mathcal M}(z_0):
        L_{G,\mathcal M}(z_0)[v]=0
    \right\}
    =
    \mathcal T_{\mathcal M}(z_0)
    \cap
    \ker(DG_{h_{z_0}}).
\]
Substitution into the preceding identity yields
\[
    r_{G,\mathcal M}(z_0)
    =
    r_{\mathcal M}(z_0)
    -
    \dim
    \left\{
        \mathcal T_{\mathcal M}(z_0)
        \cap
        \ker(DG_{h_{z_0}})
    \right\}.
\]

We next prove part~(ii).
For a vertically invariant representation satisfying the stated quotient-space
regularity conditions, $G$ induces a map $\widetilde G$ on the quotient
log-hazard space such that, locally,
$
    \Psi_G
    =
    \widetilde G
    \circ
    \pi
    \circ
    \ell.
$
The chain rule therefore gives
\[
    D\Psi_G(z_0)
    =
    D\widetilde G_{\pi(\ell_{z_0})}
    \circ
    D(\pi\circ\ell)_{z_0}.
\]
Hence
$
    \operatorname{rank}\{D\Psi_G(z_0)\}
    \leq
    \operatorname{rank}\{D(\pi\circ\ell)_{z_0}\}.
$
By Definition~\ref{def:temporal_tangent_rank},
$
    \operatorname{Im}\{D(\pi\circ\ell)_{z_0}\}
    =
    \bar{\mathcal T}_{\mathcal M}(z_0),
$
so
$
    \operatorname{rank}\{D(\pi\circ\ell)_{z_0}\}
    =
    r_{\mathrm{temp},\mathcal M}(z_0).
$
Therefore,
$
    r_{G,\mathcal M}(z_0)
    \leq
    r_{\mathrm{temp},\mathcal M}(z_0),
$
which proves part~(ii).
\end{proof}

\section{Proof of Proposition~\ref{prop:rank_reparameterization}}
\label{proof:rank_reparameterization}

\begin{proof}
Let
$
    J_\phi
    =
    D\phi_{\vartheta_0}.
$
Since $\phi$ is a local $C^1$ diffeomorphism at $\vartheta_0$,
$J_\phi$ is invertible. By the chain rule,
$
    D\widetilde h_{\vartheta_0}
    =
    Dh_{z_0}\circ J_\phi.
$
Hence
$
    \operatorname{Im}(D\widetilde h_{\vartheta_0})
    =
    \operatorname{Im}(Dh_{z_0}),
$
because $J_\phi$ is onto. Therefore the two parameterizations generate the
same local hazard tangent space,
\[
    \widetilde{\mathcal T}_{\mathcal M}(\vartheta_0)
    =
    \mathcal T_{\mathcal M}(z_0),
\]
and consequently
$
    \widetilde r_{\mathcal M}(\vartheta_0)
    =
    r_{\mathcal M}(z_0).
$

For a model-path regular representation $G$, write
$
    \Psi_G(z)=G(h_z),
$
and
$
    \widetilde\Psi_G(\vartheta)
    =
    G(\widetilde h_\vartheta)
    =
    \Psi_G\{\phi(\vartheta)\}.
$
Again by the chain rule,
$
    D\widetilde\Psi_G(\vartheta_0)
    =
    D\Psi_G(z_0)\circ J_\phi.
$
Since $J_\phi$ is invertible,
\[
    \operatorname{rank}
    D\widetilde\Psi_G(\vartheta_0)
    =
    \operatorname{rank}
    D\Psi_G(z_0),
\]
which gives
$
    \widetilde r_{G,\mathcal M}(\vartheta_0)
    =
    r_{G,\mathcal M}(z_0).
$

For the temporal rank, let
$
    \ell_z=\log h_z,
$
and
$
    \widetilde\ell_\vartheta
    =
    \ell_{\phi(\vartheta)}.
$
Whenever the corresponding temporal tangent spaces are well defined,
$
    D(\pi\circ\widetilde\ell)_{\vartheta_0}
    =
    D(\pi\circ\ell)_{z_0}\circ J_\phi.
$
The invertibility of $J_\phi$ therefore implies
$
    \widetilde{\bar{\mathcal T}}_{\mathcal M}(\vartheta_0)
    =
    \bar{\mathcal T}_{\mathcal M}(z_0),
$
and hence
$
    \widetilde r_{\mathrm{temp},\mathcal M}(\vartheta_0)
    =
    r_{\mathrm{temp},\mathcal M}(z_0).
$

It remains to consider first-order compatibility. For any
$a\in\mathbb R^q$,
\[
    D\widetilde h_{\vartheta_0}[a]
    =
    Dh_{z_0}[J_\phi a],
    \qquad
    D\widetilde\Psi_G(\vartheta_0)[a]
    =
    D\Psi_G(z_0)[J_\phi a].
\]
Thus, if
$
    Dh_{z_0}[b]=0
    \Longrightarrow
    D\Psi_G(z_0)[b]=0
$
for every $b\in\mathbb R^q$, the same implication holds under the
$\vartheta$-parameterization. The converse follows by applying the same
argument with $J_\phi^{-1}$. Hence the first-order compatibility condition is
invariant under the reparameterization.

Finally, since both $r_{\mathcal M}$ and $r_{G,\mathcal M}$ are invariant,
their difference is invariant, and the equality
$
    r_{G,\mathcal M}(z_0)
    =
    r_{\mathcal M}(z_0)
$
holds under one parameterization if and only if it holds under the other.
Therefore local functional rank loss and local completeness are also invariant.
\end{proof}

\section{Proof of Proposition~\ref{prop:covariate_induced_rank_bound}}
\label{proof:covariate_induced_rank}

\begin{proof}
By the chain rule,
$
    \left.
    D_x h_{\eta(x)}
    \right|_{x=x_0}
    =
    Dh_{z_0}\circ D\eta_{x_0},
$
and, since
$
    \Phi_G(x)
    =
    \Psi_G\{\eta(x)\},
$
we likewise have
$
    D\Phi_G(x_0)
    =
    D\Psi_G(z_0)\circ D\eta_{x_0}.
$

For linear maps $A$ and $B$,
$
    \operatorname{rank}(A\circ B)
    \leq
    \min\{\operatorname{rank}(A),\operatorname{rank}(B)\}.
$
Since
$
    \operatorname{rank}(Dh_{z_0})
    =
    r_{\mathcal M}(z_0),
$
it follows that
$
    r_X(x_0)
    \leq
    \min
    \left\{
        r_{\mathcal M}(z_0),
        \operatorname{rank}(D\eta_{x_0})
    \right\}.
$
Similarly, because
$
    \operatorname{rank}\{D\Psi_G(z_0)\}
    =
    r_{G,\mathcal M}(z_0),
$
we obtain
$
    r_{G,X}(x_0)
    \leq
    \min
    \left\{
        r_{G,\mathcal M}(z_0),
        \operatorname{rank}(D\eta_{x_0})
    \right\}.
$

Finally, suppose that the first-order compatibility condition holds at $z_0$.
By Theorem~\ref{thm:functional_rank_bound}(i),
\[
    D\Psi_G(z_0)
    =
    L_{G,\mathcal M}(z_0)\circ Dh_{z_0}.
\]
Therefore
$
    D\Phi_G(x_0)
    =
    L_{G,\mathcal M}(z_0)
    \circ
    Dh_{z_0}
    \circ
    D\eta_{x_0}.
$
Hence
\[
    \operatorname{rank}\{D\Phi_G(x_0)\}
    \leq
    \operatorname{rank}
    \left\{
        Dh_{z_0}\circ D\eta_{x_0}
    \right\},
\]
which is precisely
$
    r_{G,X}(x_0)
    \leq
    r_X(x_0).
$
\end{proof}

\section{Proof of Proposition~\ref{prop:pw_rank}}
\label{app:powerwarp_rank}

\begin{proof}
At the identity deformation state, the two log-hazard tangent directions are
$
    \dot\ell_\lambda(t)
    =
    -1-t\ell_0'(t),
$
and
$
    \dot\ell_\zeta(t)
    =
    1+\log t+t\log t\,\ell_0'(t).
$
By assumption, their quotient projections $\pi(\dot\ell_\lambda)$ and $\pi(\dot\ell_\zeta)$ are linearly independent in $\mathbb A_T/\mathbb V$.
Hence the projected temporal tangent space is two-dimensional and $r_{\mathrm{temp},\mathcal M_{\mathrm{PW}}}(0,0) = 2$.

Linear independence after projection implies linear independence of $\dot\ell_\lambda$ and $\dot\ell_\zeta$ in the original log-hazard space.
The corresponding hazard-scale tangent directions are $\dot h_\lambda = h_0\dot\ell_\lambda$, $\dot h_\zeta = h_0\dot\ell_\zeta$.
Since $h_0>0$ on $[\varepsilon,T]$, multiplication by $h_0$ is injective, so these two hazard-scale tangent directions are also linearly independent.
Therefore $r_{\mathcal M_{\mathrm{PW}}}(0,0) = 2$.
\end{proof}

\section{Proof of Proposition~\ref{prop:llp_peak_completeness}}
\label{proof:llp_peak_completeness}

\begin{proof}
Write
$
    k=\alpha\delta>1,
$
$    
    y=\frac{t}{\mu}.
$
With $\alpha$ fixed, the LLP hazard can be written as
\[
    h(t)
    =
    \frac{k}{\mu}
    \frac{y^{k-1}}{1+y^k}.
\]
Differentiating with respect to $y$ gives
\[
    \frac{d}{dy}
    \left\{
        \frac{y^{k-1}}{1+y^k}
    \right\}
    =
    \frac{
        y^{k-2}\{(k-1)-y^k\}
    }{
        (1+y^k)^2
    }.
\]
Since $k>1$, the derivative changes sign exactly once, at
$
    y_{\mathrm{peak}}
    =
    (k-1)^{1/k}.
$
Hence the hazard has a unique interior peak, with
$
    t_{\mathrm{peak}}
    =
    \mu(k-1)^{1/k},
$
$    
    h_{\mathrm{peak}}
    = (k-1)^{(k-1)/k}/\mu.
$
Moreover, differentiating once more and evaluating at the peak gives
\[
    h''(t_{\mathrm{peak}})
    =
    -\frac{y_{\mathrm{peak}}^{\,2k-3}}{\mu^3}
    <0,
\]
so the peak is nondegenerate.

The peak pair satisfies
$
    t_{\mathrm{peak}}h_{\mathrm{peak}}
    =
    k-1,
$
and therefore determines $(\mu,\delta)$ uniquely through
\[
    \delta
    =
    \frac{1+t_{\mathrm{peak}}h_{\mathrm{peak}}}{\alpha},
    \qquad
    \mu
    =
    \frac{t_{\mathrm{peak}}}
    {
        (t_{\mathrm{peak}}h_{\mathrm{peak}})
        ^{1/\{1+t_{\mathrm{peak}}h_{\mathrm{peak}}\}}
    }.
\]
Both the forward map and its inverse are smooth on their respective domains.
Thus
\[
    (\mu,\delta)
    \longmapsto
    \{t_{\mathrm{peak}},h_{\mathrm{peak}}\}
\]
is smoothly invertible on $\mathcal D_\alpha$, and its Jacobian has rank two
throughout this domain. Consequently,
$
    r_{G_{\mathrm{peak}},\mathcal M}=2.
$

For any fixed $(\mu,\delta)\in\mathcal D_\alpha$, one may choose a compact
time interval containing $t_{\mathrm{peak}}$ in its interior; by continuity of
the peak map, this remains true in a neighbourhood of $(\mu,\delta)$.
The smoothness and nondegeneracy conditions above therefore allow
Proposition~\ref{prop:peak_functional_regularity} to be applied locally, so the
first-order compatibility condition holds.

Since the fixed-$\alpha$ LLP hazard family is parameterized by the two
coordinates $(\mu,\delta)$,
$
    r_{\mathcal M}\leq 2.
$
By Theorem~\ref{thm:functional_rank_bound}(i),
$
    2
    =
    r_{G_{\mathrm{peak}},\mathcal M}
    \leq
    r_{\mathcal M}
    \leq
    2.
$
Hence
$
    r_{G_{\mathrm{peak}},\mathcal M}
    =
    r_{\mathcal M}
    =
    2,
$
and $G_{\mathrm{peak}}$ is locally complete for the fixed-$\alpha$ LLP hazard
family throughout $\mathcal D_\alpha$.
\end{proof}

\section{Proof of Proposition~\ref{prop:identifiability}}
\label{app:identifiability_proof}

\begin{proof}
Suppose that two parameter values
$
\theta
=
(\alpha,\beta_0,\boldsymbol{\beta}^{\top},
\boldsymbol{\gamma}^{\top})^{\top}
$
and
$
\theta^\star
=
(\alpha^\star,\beta_0^\star,
\boldsymbol{\beta}^{\star\top},
\boldsymbol{\gamma}^{\star\top})^{\top}
$
generate the same conditional hazard function, that is,
$
h(t\mid\mathbf{x};\theta)
=
h(t\mid\mathbf{x};\theta^\star)
$
for all \(t>0\) and all \(\mathbf{x}\in\mathcal X\).
Since
$
S(t\mid\mathbf{x})
=
\exp\left\{
-\int_0^t h(u\mid\mathbf{x})\,du
\right\},
$
equality of the conditional hazards implies equality of the corresponding conditional survival functions:
$
S(t\mid\mathbf{x};\theta)
=
S(t\mid\mathbf{x};\theta^\star)
$
for all \(t>0\) and all \(\mathbf{x}\in\mathcal X\).
Using Equation~\eqref{eq:llpw_survival}, we obtain
\[
\left\{
\frac{t}{\mu(\mathbf{x})}
\right\}^{\alpha\delta(\mathbf{x})}
=
\left\{
\frac{t}{\mu^\star(\mathbf{x})}
\right\}^{\alpha^\star\delta^\star(\mathbf{x})}.
\]
Taking logarithms gives
$
\alpha\delta(\mathbf{x})
\left\{
\log t-\log\mu(\mathbf{x})
\right\}
=
\alpha^\star\delta^\star(\mathbf{x})
\left\{
\log t-\log\mu^\star(\mathbf{x})
\right\}.
$
Substituting Equations~\eqref{eq:scaling_component} and~\eqref{eq:deformation_component} yields
\[
\alpha
\exp\left(\mathbf{x}^{\top}\boldsymbol{\gamma}\right)
\left(
\log t-\beta_0-\mathbf{x}^{\top}\boldsymbol{\beta}
\right)
=
\alpha^\star
\exp\left(\mathbf{x}^{\top}\boldsymbol{\gamma}^\star\right)
\left(
\log t-\beta_0^\star
-\mathbf{x}^{\top}\boldsymbol{\beta}^\star
\right).
\]
For each fixed \(\mathbf{x}\in\mathcal X\), both sides are affine functions of \(\log t\), and the equality holds for all \(t>0\), equivalently for all \(\log t\in\mathbb R\). Their coefficients of \(\log t\) must therefore be equal:
\[
\alpha
\exp\left(\mathbf{x}^{\top}\boldsymbol{\gamma}\right)
=
\alpha^\star
\exp\left(\mathbf{x}^{\top}\boldsymbol{\gamma}^\star\right),
\qquad
\mathbf{x}\in\mathcal X.
\]
Taking logarithms gives
$
\log\alpha-\log\alpha^\star
+
\mathbf{x}^{\top}
\left(
\boldsymbol{\gamma}
-
\boldsymbol{\gamma}^\star
\right)
=
0.
$
This is an affine function of \(\mathbf{x}\) that vanishes on \(\mathcal X\). 
Since
$
\operatorname{aff}(\mathcal X)=\mathbb R^p,
$
the only affine function vanishing on \(\mathcal X\) is the zero function.
Consequently,
$
\log\alpha-\log\alpha^\star=0,
$
$
\boldsymbol{\gamma}
=
\boldsymbol{\gamma}^\star,
$
and hence $\alpha=\alpha^\star$.

Substituting these identities into the preceding equality gives
\[
\beta_0+\mathbf{x}^{\top}\boldsymbol{\beta}
=
\beta_0^\star
+
\mathbf{x}^{\top}\boldsymbol{\beta}^\star,
\qquad
\mathbf{x}\in\mathcal X.
\]

Again, this is an affine function that vanishes on \(\mathcal X\).
The condition
$
\operatorname{aff}(\mathcal X)=\mathbb R^p
$
therefore implies
$
\beta_0=\beta_0^\star,
$
$
\boldsymbol{\beta}=\boldsymbol{\beta}^\star.
$
Thus, $\theta=\theta^\star,$ and the model is identifiable.
\end{proof}

\section{Proof of Corollary~\ref{cor:functional_covariance_rank}}
\label{proof:functional_covariance_rank}
\begin{proof}
Let $A=\dot\Psi_G(\theta_0;x)$.
Since $\Sigma_\theta$ is positive definite, it admits an invertible square root $\Sigma_\theta^{1/2}$, and
$
    V_G(x)
    =
    (A\Sigma_\theta^{1/2})
    (A\Sigma_\theta^{1/2})^\top.
$
Hence
$
    \operatorname{rank}\{V_G(x)\}
    =
    \operatorname{rank}(A\Sigma_\theta^{1/2})
    =
    \operatorname{rank}(A),
$
because $\Sigma_\theta^{1/2}$ is invertible.
The second statement follows from the definition of $r_{G,\mathcal M}(\theta_0)$.
\end{proof}

\section{Proof of Proposition~\ref{prop:llp_uniform_hazard_recovery}}
\label{proof:llp_uniform_hazard_recovery}

Since $\mathbb T$, $\mathcal X$, and $\Theta_0$ are compact and the hazard map is continuous on $\mathbb T\times\mathcal X\times\Theta_0$, it is uniformly continuous on this set.
Because $\Theta_0$ contains an open neighbourhood of $\theta_0$, there exists $r>0$ such that
$$
    \{\theta:\|\theta-\theta_0\|<r\}\subset \Theta_0.
$$
For every $\eta>0$, uniform continuity implies that there exists $\delta\in(0,r)$ such that, whenever $\|\theta-\theta_0\|<\delta$,
$$
    \sup_{x\in\mathcal X}
    \sup_{t\in\mathbb T}
    \left|
        h(t\mid x;\theta)
        -
        h(t\mid x;\theta_0)
    \right|
    <\eta.
$$
Therefore,
$$
\left\{
    \sup_{x\in\mathcal X}
    \sup_{t\in\mathbb T}
    \left|
        h(t\mid x;\widehat\theta_n)
        -
        h(t\mid x;\theta_0)
    \right|
    \ge \eta
\right\}
\subseteq
\left\{
    \|\widehat\theta_n-\theta_0\|\ge \delta
\right\}.
$$
Taking probabilities and using $\widehat\theta_n\xrightarrow{p}\theta_0$ gives
$$
P\left(
    \sup_{x\in\mathcal X}
    \sup_{t\in\mathbb T}
    \left|
        h(t\mid x;\widehat\theta_n)
        -
        h(t\mid x;\theta_0)
    \right|
    \ge \eta
\right)
\to 0.
$$
This proves the claimed uniform hazard recovery.

\section{List of Hazard Functionals}
\label{app:list_hazard_functionals}

This appendix collects the hazard functionals used in the theoretical development, simulation studies, and applied risk-profile analysis.
All functionals are defined on a compact time domain $\mathcal T=[\varepsilon,T_0]$, with $\varepsilon>0$.
Throughout, $h\in\mathcal H$ denotes a hazard trajectory on $\mathcal T$.

\begin{enumerate}
    \item \textbf{Window-specific cumulative hazard.}
    For $\varepsilon\leq a<b\leq T_0$,
    $
    H_{a,b}(h)
    =
    \int_a^b h(u)\,du.
    $
    In particular, the cumulative hazard up to a prediction horizon $\tau\leq T_0$ is
    $
    H_\tau(h)
    =
    \int_\varepsilon^\tau h(u)\,du.
    $

    \item \textbf{Fixed-horizon event risk.}
    The event risk induced by the cumulative hazard up to horizon $\tau$ is
    $
    R_\tau(h)
    =
    1-\exp\{-H_\tau(h)\}
    =
    1-\exp\left\{-\int_\varepsilon^\tau h(u)\,du\right\}.
    $
    This includes the fixed-horizon risks $R_2(h)$ and $R_4(h)$ used in the simulation studies.

    \item \textbf{Peak timing.}
    The peak time of the hazard trajectory is
    $
    t_{\mathrm{peak}}(h)
    =
    \arg\max_{t\in[\varepsilon,T_0]} h(t).
    $
    When the maximizer is not unique, the earliest maximizer is used.

    \item \textbf{Peak intensity.}
    The peak height, or maximum instantaneous hazard, is
    $
    h_{\mathrm{peak}}(h)
    =
    \max_{t\in[\varepsilon,T_0]} h(t).
    $

    \item \textbf{High-risk width.}
    For a threshold fraction $0<\rho<1$, the high-risk width is
    \[
    w_\rho(h)
    =
    \lambda\left\{
    t\in[\varepsilon,T_0]:
    h(t)\geq \rho h_{\mathrm{peak}}(h)
    \right\},
    \]
    where $\lambda(\cdot)$ denotes Lebesgue measure.
    The half-peak width $w_{0.5}(h)$ is used as the default persistence descriptor.

    \item \textbf{Early hazard mass.}
    For an early-risk cutoff $\tau_E$ satisfying $\varepsilon<\tau_E<T_0$,
    $
    A_E(h)
    =
    \int_\varepsilon^{\tau_E}h(u)\,du.
    $

    \item \textbf{Late hazard mass.}
    For a late-risk cutoff $\tau_L$ satisfying $\varepsilon<\tau_L<T_0$,
    $
    A_L(h)
    =
    \int_{\tau_L}^{T_0}h(u)\,du.
    $

    \item \textbf{Tail ratio.}
    For $\varepsilon<\tau_E\leq \tau_L<T_0$, the tail ratio is
    \[
    q_{\mathrm{tail},\tau_E,\tau_L}(h)
    =
    \frac{A_L(h)}{A_E(h)}
    =
    \frac{
    \int_{\tau_L}^{T_0}h(u)\,du
    }{
    \int_\varepsilon^{\tau_E}h(u)\,du
    }.
    \]
    The log-tail ratio is
    $
    \log q_{\mathrm{tail},\tau_E,\tau_L}(h),
    $
    which is often used for numerical stability and interpretability.
    A midpoint split is obtained as the special case $\tau_E=\tau_L=T_0/2$.



\end{enumerate}

\end{document}